\pdfoutput=1
\documentclass[sigconf]{acmart}

\usepackage[english]{babel}
\usepackage{blindtext}
\usepackage{float}
\usepackage{booktabs}

 \usepackage[all=normal,paragraphs=tight]{savetrees}

\setcopyright{none}

\usepackage{subcaption}

\graphicspath{ {./figures/} }

\usepackage[htt]{hyphenat}  
\usepackage{mathtools}
\DeclarePairedDelimiter{\floor}{\lfloor}{\rfloor}

\usepackage{multirow}

\begin{document}

 \title{A Study of Feasibility and Diversity of Web Audio Fingerprints}

\author{Shekhar Chalise}
   \affiliation{
   \institution{University of New Orleans}
   \city{New Orleans} 
   \state{LA} 
  }
\email{shalise@my.uno.edu}

\author{Phani Vadrevu}
   \affiliation{
   \institution{University of New Orleans}
   \city{New Orleans} 
   \state{LA} 
  }
\email{phani@cs.uno.edu}


\begin{abstract}
Prior measurement studies on browser fingerprinting have unfortunately largely excluded Web Audio API-based fingerprinting in their analysis. We aim to address this issue by conducting \emph{the first systematic study of effectiveness of web audio fingerprinting} mechanisms. We focus on studying the \emph{feasibility} (which includes \emph{stability} and \emph{timing} aspects) and \emph{diversity} properties of web audio fingerprinting.  Along with 3 known audio fingerprinting vectors, we designed and implemented 4 new audio fingerprint vectors that work by obtaining FFTs of waveforms generated via different methods. Our study used MTurk and other social media platforms, to collect and analyze audio fingerprints from 2093 web users.

Our results present new insights into the nature of Web Audio fingerprints. First, we show that audio fingeprinting vectors, unlike other prior vectors, reveal an apparent fickleness with some users' browsers giving away differing fingerprints in repeated attempts. However, we show that it is possible to devise a graph-based analysis mechanism to collectively consider all the different fingerprints left by users' browsers and thus craft a highly stable fingerprinting mechanism. Our analysis also shows that it is possible to do this in a timely fashion with each vector taking only about 0.14 seconds of time on average. 

Next, we investigate the diversity of audio fingerprints and compare this with prior fingerprinting techniques.  Our results show that audio fingerprints are much less diverse than other vectors with only 95 distinct fingerprints among 2093 users. At the same time, further analysis shows that web audio fingerprinting can potentially bring considerable additive value (in terms of entropy) to existing fingerprinting mechanisms. We also show that our results contradict the current security and privacy recommendations provided by W3C regarding audio fingerprinting. Overall, our systematic study allows browser developers to gauge the degree of privacy invasion presented by audio fingerprinting thus helping them take a more informed stance when designing privacy protection features in the future.

\end{abstract}
\maketitle

\section{Motivation}
\label{sec:introduction}

Browser fingerprinting presents a grave threat to the privacy of internet users as it allows user tracking even in private browsing modes. The recent advent of HTML5 and advanced web APIs has tremendously increased the fingerprintable surface area of web browsers. As a result, security and privacy researchers have extensively focused on measuring and tracking the evolution of browser fingerprints obtained by using APIs such as Canvas and WebGL~\cite{LaperdrixRB16,CaoLW17,GomezLB18,VastelLRR18} in order to quantify the scope of the problem. However, despite being used in the wild since 2016~\cite{EnglehardtN16,Anupam18}, Web Audio API-based fingerprinting has remained a notable absence in such large-scale fingerprint measurement works. In particular, to our knowledge, there exists no prior work that systematically measures the effectiveness of various Web Audio-based fingerprinting techniques and compares them with existing fingerprinting techniques to gauge their relative importance. In this work, we attempt to fill this important knowledge gap.

The Web Audio API is a powerful system that enables websites to dynamically edit audio and perform complicated mixing operations such as creating spatial effects and audio visualizations as well as  mixing different audio sources. In the absence of a dedicated study on the effectiveness of Web Audio-based browser fingerprints, browser developers have been left to speculate about how essential defenses are for audio fingerprinting attacks. This resulted in varying levels of protections across different web browsers. For example, the Brave Browser offers default randomization-based audio fingerprinting defenses~\cite{brave_audio_fp_defense}. This solution adds small random modifications (< 1\%) to the amplitudes of audio signals generated from Web Audio APIs as proposed and implemented in~\cite{LaperdrixBM17} . Thus, the browser produces a slightly different signal each time the fingerprinting code is run and prevents identification of the users. However, with such fingerprinting solutions, web browsers risk exposing users to computational overhead as well as compatibility issues with web sites that have legitimate use cases for the targeted APIs~\cite{brave_canvas_compat, brave_canvas_compat2}. On the other hand, other browsers such as Chrome and Firefox do not agree with Brave's randomization approach with questions arising about the seriousness of the fingerprinting surface exposed by Web Audio APIs~\cite{rtoy_2020, padenot_2020}. This is also presented in the World Wide Web Consoritum's (W3C's) Wed Audio API standards document which states that web audio fingerprinting ``merely allows deduction of information already readily available by easier means (User Agent string)''~\cite{wwwc_2021}. One of our goals with this work is to be able to collect and analyze audio fingerprinting data to confirm or refute this assertion. Overall, we believe that our work will be beneficial to the browser developer community to take a more informed stance towards audio fingerprinting defenses. 

We performed our experiments by recruiting 2093 volunteers from 57 countries all over the world with the help of Amazon's MTurk platform as well as leveraging our social circles. Our study's participants spent a total estimated time of about 108 hours on our fingerprinting web site as part of this study. Our website contained fingerprinting code for 3 previously known audio fingerprinting techniques~\cite{openwpm_afp}, as well as 4 new audio fingerprint vectors that we devised to rely on Finite Fourier Transforms (FFTs) of modulated wave forms generated with the help of various Web Audio APIs.  With this setup, we conducted the first systematic study of effectiveness of Web Audio-based browser fingerprinting vectors. Our study's major contributions can summarized as follows:


\begin{enumerate}
\item \emph{Web Audio Fingerprinting Vectors.} We designed and implemented 4 new audio fingerprinting vectors that made use of Fast Fourier transformations of modulated custom waveforms.
\item \emph{User Study.} We collected basic web audio configuration information, 7 Web Audio API-based fingerprints as well as multiple other well known browser fingerprints via an elaborate worldwide user study involving 2093 users.
\item \emph{Feasibility Analysis.} We designed a graph-based fingerprint mechanism to collate the multiple audio fingerprints associated with each user. Using this mechanism, we demonstrated that Web Audio APIs can be utilized to yield a stable browser fingerprinting system.  
\item \emph{Diversity and Effectiveness.} We presented diversity measures of audio fingerprints. We also showed the relative effectiveness of these fingerprints in comparison to other browser fingerprinting vectors such as Canvas, Font and \texttt{User-Agent}-based fingerprinting to help future browser developers to take informed design decisions regarding privacy protection.
\end{enumerate}

We also state that in order to support research in web privacy protection, we will share our fingerprinting code, analysis code as well as the anonymized fingerprinting data sets and results that we obtained in our research with all vetted security and privacy researchers from academia as well as industry. This can be particularly impactful given that browsers developers focusing on privacy measures are currently actively relying on such measurement data to derive design decisions~\cite{rtoy_2020_metrics, rtoy_2018_metrics}.

\subsubsection*{Paper Roadmap.} The rest of the paper is organized as follows. Section~\ref{sec:sys_desc} covers our system description where prior web audio fingerprinting vectors are first discussed (Section~\ref{ssec:prior_fp}) and then the new vectors that we propose to study are covered (Section~\ref{ssec:new_fp}). This is followed by description of the details of our fingerprinting code (Section~\ref{ssec:exp_setup}) and then the demographics of the participants in our study (Section~\ref{ssec:participants}). The feasibility analysis begins with a preliminary analysis of user study data (Section~\ref{ssec:stability_prelim}) followed by a proposal for a fingerprint collation system (Section~\ref{ssec:graph_approach}). This system is then put to use for stability analysis of the data (Section~\ref{ssec:stability_analysis}) after which a timing analysis (Section~\ref{ssec:timing}) is also done confirming the feasibility of audio fingerprints. After this, in Section~\ref{sec:diversity}, the paper dives into diversity analysis of audio fingerprints (both standalone and relative) with special focus on comparison with Canvas and \texttt{User-Agent} fingerprinting mechanisms to provide guidance to browser developers. Potential criticism and limitations are discussed in Section~\ref{sec:discussion}. Related work is covered in Section~\ref{sec:related_work} before concluding the paper in Section~\ref{sec:conclusion}.

\section{System Description}
\label{sec:sys_desc}

In this section, we will furnish the details of our experimental setup to study the effectiveness of Web Audio-based fingerprinting in web browsers. The Web Audio API was first introduced by Google in 2011~\cite{rogers_2011} in order to enable synthesis and processing of audio on the web with support for fine-grained timing controls, real-time sound effects as well as complex visualizations. The use of the API involves the creation of an ``Audio Graph'' which is a directed graph built by the users to enable arbitrarily complex audio modifications. The atomic components of this graph are the ``Audio Nodes'' which can represent any audio modules such as audio sources (files, synthesizers etc.), destinations (speakers, offline buffers etc.), modifiers and analyzers. 


\subsection{Prior Audio Fingerprinting Vectors}
\label{ssec:prior_fp}
We will now describe some prior audio fingerprinting vectors that have been discovered in the wild whose effectiveness we plan to systematically study in this work.

\subsubsection*{Dynamics Compressor (DC)}

\begin{figure}[h]
    \centering
    \includegraphics[width=1.0\columnwidth]{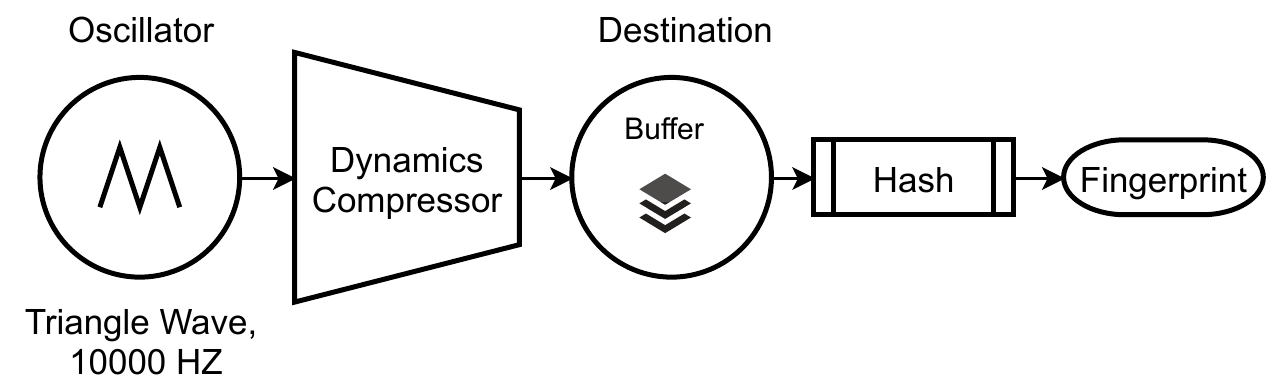}
    \caption{Dynamics Compressor (DC) Method}
    \label{fig:DynamicsCompressor}
\end{figure}

Englehardt et al. have discovered two audio fingerprinting methods being used in the wild in their prior work~\cite{EnglehardtN16}. One of these is the Dynamics Compressor (DC) method whose audio graph is depicted in Figure~\ref{fig:DynamicsCompressor}. The method simply involves the use of an~\texttt{OscillatorNode} to create a periodic audio waveform in a specific shape (such as a \texttt{triangle}) and feeding it to a DC Node (\texttt{DynamicsCompressorNode}). DC is a often used method in muscial production to lower the volume of the loudest parts in the audio. This allows to reduce distortion and clipping effects that commonly exist in recorded audio samples. This fingerprinting vector's main intuition is that there might exist small identifiable differences in the way dynamics compression is done in different audio hardware/software stacks of different users. Hence, to capitalize on this, this method directs the output of DC to an offline buffer which is then sent to a hash function to produce the final fingerprint.

\subsubsection*{Fast Fourier Transform (FFT)}
\begin{figure}[h]
    \centering
    \includegraphics[width=1.0\columnwidth]{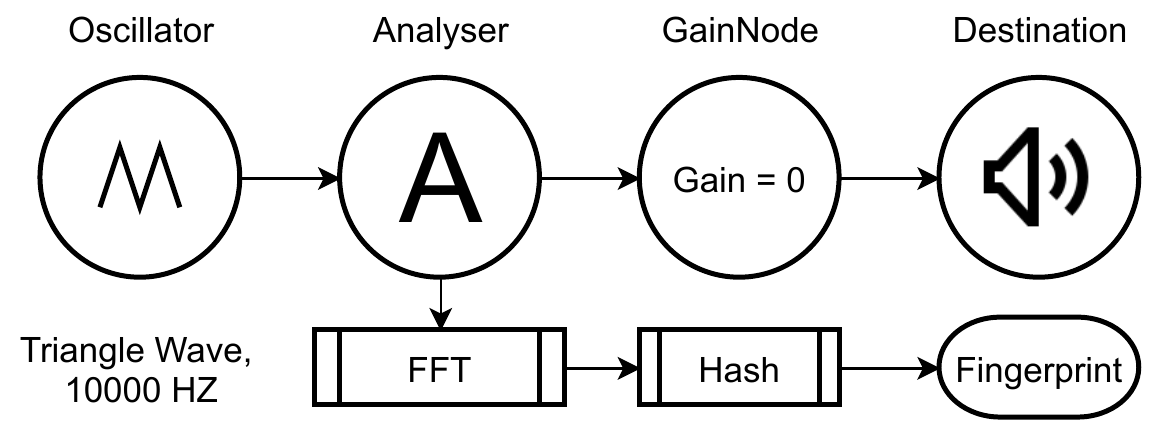}
    \caption{Fast Fourier Transform (FFT) Method}
    \label{fig:fft}
\end{figure}

The second audio fingerprinting method discovered in~\cite{EnglehardtN16} is depicted in Figure~\ref{fig:fft} where the intuition is to make use of tiny but characteristic differences that might exist in the Fast Fourier Transformation (FFT) calculations performed by the web browsers when requested to transform a simple audio signal from time domain to frequency domain. As seen in the figure, this is accomplished with the help of an~\texttt{AnalyserNode} and a \texttt{ScriptProcessorNode}  after which the FFT output is sent to a hash function to produce the final output. Note that this fingerprinting method uses an alternate method of ``silencing the fingerprinting audio'' by sending it to a \texttt{GainNode} whose gain (volume) is set to zero before sending the output to an online destination such as the computer speakers.

\subsubsection*{Hybrid (DC + FFT)}
\begin{figure}[h]
    \centering
    \includegraphics[width=1.0\columnwidth]{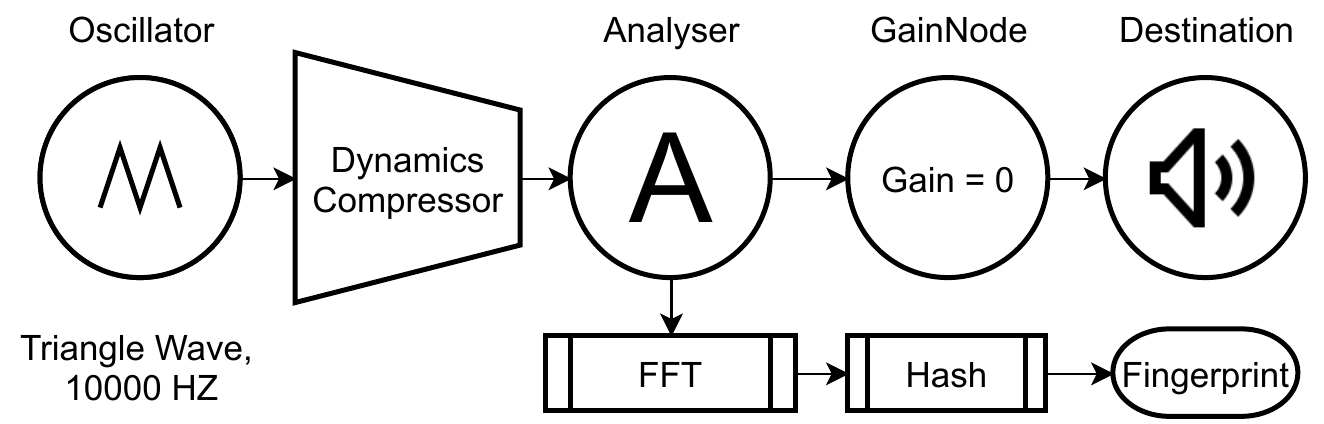}
    \caption{Hybrid (DC + FFT) Method}
    \label{fig:Hybrid}
\end{figure}

The authors of \cite{EnglehardtN16} also developed another audio fingerprinting method that simply combines both DC and FFT in an attempt to increase the amount of ``fingerprintability''~\cite{openwpm_afp} as is depicted in Figure~\ref{fig:Hybrid}. We also included this in our work to study its effectiveness as it represents the widest audio fingerprinting vector proposed thus far. The authors refer to this method as a ``hybrid'' audio fingerprinting vector and we will use this same notation in the rest of this paper.

We obtained the code samples for all these three vectors from~\cite{openwpm_afp} used them as part of our audio fingerprinting array. 

\subsection{New Audio Fingerprinting Vectors}
\label{ssec:new_fp}

Along with studying the effectiveness of known audio fingerprinting vectors, we also wanted to see if it is possible to improve these vectors in order to increase their ``fingerprintability'' of audio software/hardware stack. For this, we created 4 new vectors by extending the hybrid (DC + FFT) vector. In all the vectors, we attempted to create more complicated signals so as to increase diversity in fingerprints. We describe these below.

\subsubsection*{Merged Signals} 
\begin{figure}
    \centering
    \includegraphics[width=1.0\columnwidth]{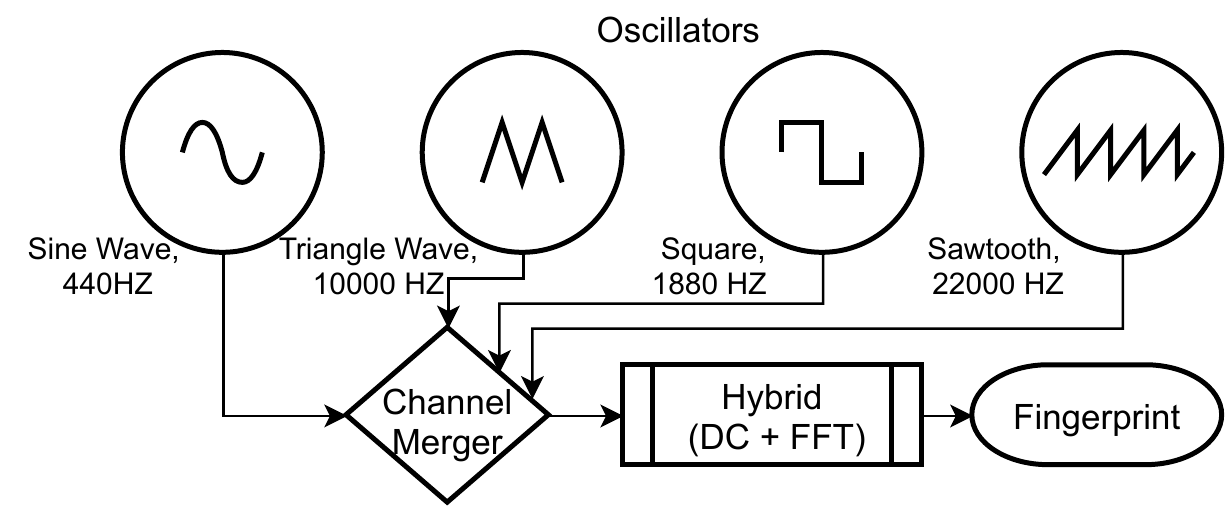}
    \caption{Merged Signals Method}
    \label{fig:ChannelMergeHybrid}
\end{figure}

Our first idea in extending the earlier hybrid vector is to simply use multiple signals instead of the single \texttt{triangle} signal. This is depicted in Figure~\ref{fig:ChannelMergeHybrid}. Our main idea was to check if using other shapes of the waves could potentially increase the diversity of fingerprints. For this, we used all the four shapes of waves supported by~\texttt{OscillatorNode} (generated in different frequencies). We then merged them together using~\texttt{ChannelMergerNode} which is usually used to combine mono audio inputs (such as L,R,C etc) into a single output channel. The rest of the fingerprinting mechansim is the same as that of the hybrid method.

\subsubsection*{Custom Signal} For our second vector, we used the `\texttt{custom}' wave shape type supported by \texttt{OscillatorNode} which allowed us to define our own wave shape. We used an array of 12 real and imaginary values to define this periodic signal with real values randomly selected between 0 and 1 and imaginary values alternating between 0 and $\pi$/2. It is to be noted that a `\texttt{custom}' wave type was also used as an input to a DC fingerprinting vector in~\cite{Queiroz} previously. More detailed comparison with~\cite{Queiroz} is presented in Section~\ref{sec:related_work}.


\subsubsection*{Amplitude Modulation (AM)} 
\begin{figure}[H]
    \centering
    \includegraphics[width=1.0\columnwidth]{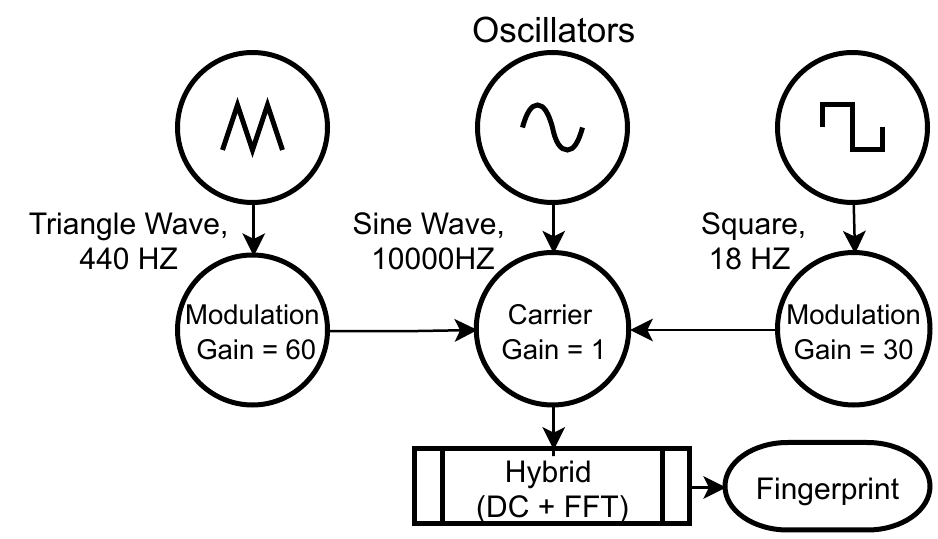}
    \caption{Amplitude Modulation Method}
    \label{fig:am}
\end{figure}

We also wanted to create an Amplitude Modulated (AM) wave signal in order to see if the process of modulation increases the fingerprint diversity. For this, as depicted in Figure~\ref{fig:am}, we generate two waves (triangle and square) and modulate them with the help of another generated sine wave as a carrier wave. 

\subsubsection*{Frequency Modulation (FM)} This final method is the same as previous AM method except that we used Frequency Modulation (FM) instead.

\subsection{Experimental Setup}
\label{ssec:exp_setup}
We set up a web page to host the fingerprinting code for all the 7 vectors we discussed above. As described in Section~\ref{sec:introduction}, our main goal is to evaluate the effectiveness of web audio fingerprinting. In order for any fingerprint to be effective, we need to measure both the \emph{feasibility} of the fingerprinting mechanism as well as \emph{diversity} of resulting fingerprints. The feasibility analysis itself can be divided into two components: \emph{stability} analysis and \emph{timing} analysis. Stability of a fingerprinting mechanism means that the same user with the same browser should result in the same fingerprint even if fingerprinted repeatedly. This is the cornerstone of any fingerprinting mechanism and it is important to verify this in our study. For this, we designed our study's web page to repeatedly run the same audio fingerprinting code multiple times. This allowed to us to collect multiple fingerprints for each vector from each participant in the study and thus evaluate the stability aspects of the various fingerprinting mechanisms we consider. 

Further, it is also important to perform timing analysis of fingerprinting mechanisms in order to ensure that the code does not take too long to run on any machine. For this, we capture compute times for each fingerprinting vector and for each iteration. Moreover, we also included other previously known fingerprinting vectors such as Canvas, Font and User-Agent to enable us to evaluate the relative effectiveness of audio fingerprinting vectors. We leveraged code from a popular browser fingerprinting project (AmIUnique) for this purpose~\cite{amiunique}. Since the stability of these other fingerprints is either pre-established by definition or has been proven experimentally~\cite{MoweryS12}, we only extract these fingerprints one time for each visit of a user to our web page.  We also used an open source fingerprinting library (FingerprintJS)~\cite{fingerprintjs} to extract basic web audio configuration information such as the default sample rate and audio channel count. However, it is to be noted that a prior work has already shown that the diversity of these vectors is too small to be used as a fingerprinting vector~\cite{LaperdrixBM17}. We still included this in the study for the sake of completeness.

For the 7 audio fingerprints, in order to decide how many times we needed to repeatedly capture each fingerprint from each user, we ran pilot experiments during which we profiled our code in commodity laptops and personal devices. We noticed that by setting the number of iterations to 30, our entire fingerprinting code ran for about 30 seconds to one minute on these different machines. Since this amount of time matched the planned time for each volunteer in our study, we used 30 as the number of iterations. As a result, our web page was set up to collect a total of 210 audio fingerprints (30 iterations, 7 vectors) from each user.

Our project's fingerprinting website was built with 5800 lines of TypeScript code using the Angular 11.0.4 framework and the Cloud Firebase database. We also wrote about 10,000 lines of Python code for all the fingerprint analysis presented in this manuscript. 

\subsection{Participants}
\label{ssec:participants}

We recruited participants for our study via Amazon's MTurk platform as well as via our social circles. We obtained an IRB exemption from our university before conducting the study. We also also presented a clear disclosure messsage to the participants informing about audio fingerprint extraction prior to the beginning of the study. For MTurk, we modeled the study request as an MTurk survey in order to make sure no particular participant participates more than once in the study. Further, we also added a filtering step where we only consider one dataset record for each IP address and user agent pair. This additional step is to ensure that only unique participants are considered in our analysis and thus prevent any accidental repeat visits from tampering our results.

Our study was conducted for 76 days during the months of March to May 2021. During this time, our website received 2605 visits. After filtering out duplicate records as mentioned above as well as pruning incomplete records, we were left with 2093 records thus indicating 2093 unique participants in our user study. We estimate the total time spent by all participants on our fingerprinting web page to be about 108 hours with each user spending an average of 2.5 minutes on our web page. We had a very diverse participant pool covering as many as 57 different countries. Among those countries, the United States, India, Brazil and Italy were the most frequent with each of them having at least 100 participants. 

From the~\texttt{User-Agent} HTTP headers, we inferred that our participants used different browsers such as Google Chrome, Mozilla Firefox as well as several Chrome-based browsers such as Microsoft Edge, Opera, Samsung Internet, Amazon Silk, Yandex and MIUI browsers. Firefox was used by about 9.6\% of the participants while the remaining 90.4\% all used Chrome-based browsers. Our study also included all major OS families such as Windows (78.5\%), Android (6.9\%), MacOS (9.4\%) and Linux (5.2\%). We excluded iOS devices from our study due to the additional technical complications introduced by Apple's explicit user action requirement for creating each audio object~\cite{ios_webaudio}. 
\section{Feasibility Analysis}
\label{sec:feasibility}
\subsection{Preliminary Analysis}
\label{ssec:stability_prelim}

\begin{table}[h]
\small
\centering
    \begin{tabular}{lccc}
    \toprule
    {\bf Vector} & {\bf Min.} & {\bf Max.} & {\bf Mean}\\
    \midrule
    DC & 1 & 1 & 1.0 \\ 
    \midrule
    FFT & 1 & 21 & 1.807 \\ 
    \midrule
    Hybrid (DC+FFT) & 1 & 18 & 2.082 \\ 
    \midrule
    Custom Signal & 1 & 18 & 2.084 \\ 
    \midrule
    Merged Signals & 1 & 21 & 2.922 \\ 
    \midrule
    AM & 1 & 26 & 4.284 \\ 
    \midrule
    FM & 1 & 24 & 4.334 \\ 
    \midrule
    \end{tabular}
    \caption{Distinct fingerprints per user among 30 iterations}
    \label{table:fp_stats}
\end{table}

As described in Section~\ref{ssec:exp_setup}, in order to gauge the feasibility of Web Audio-based fingerprinting, we first analyze the stability of the fingerprints and then inspect the timing aspects. When conducting  a preliminary analysis of the results for stability, we observed that the Web Audio API-based fingerprints have some ``fickleness'' with some users' browsers leaving more than 20 different fingerprints among the 30 iterations we make for each vector. These numbers are shown in the ``Max.'' column of Table~\ref{table:fp_stats}. This phenomenon appears unique to the Web Audio API-based fingerprinting as other HTML5 APIs abused for fingerprinting such as Canvas and WebGL~\cite{MoweryS12} have been shown to be very stable (unless there is a browser upgrade).  Among the ``Max.'' values, the Dynamics Compressor (DC) vector stands out in the table as it results in only one stable fingerprint for each of the 2093 users across all 30 iterations. All the other vectors including the Hybrid vector are showing varying number of fingerprints (of at least 18 or more) across different iterations for some of the users. As the FFT is the only difference between Hybrid and DC vectors (see Section~\ref{ssec:prior_fp}), it is likely that FFT calculations are what are causing this apparent instability in the extracted fingerprints.

Interestingly, we found that audio fingerprints extracted from Chrome-based browsers exhibit much more fickleness than Firefox. For example, less than 0.2\% of Chrome browsers (3) had only a single AM fingerprint across all 30 iterations where as more than 97.5\% of Firefox browsers (197) resulted in only a single fingerprint for the AM vector in our study. We observed similar differences for FM (0.1\% vs. 96\%) as well as Hybrid vectors (42.3\% vs. 97.5\%) too. This indicates that differences in the way web audio APIs are implemented by browser developers are likely contributing to different fingerprints across iterations. We plan to investigate this more in future after disclosure to browser developers.


\begin{figure*}
        \centering
        \begin{subfigure}[t]{0.49\textwidth}
            \includegraphics[width=\textwidth]{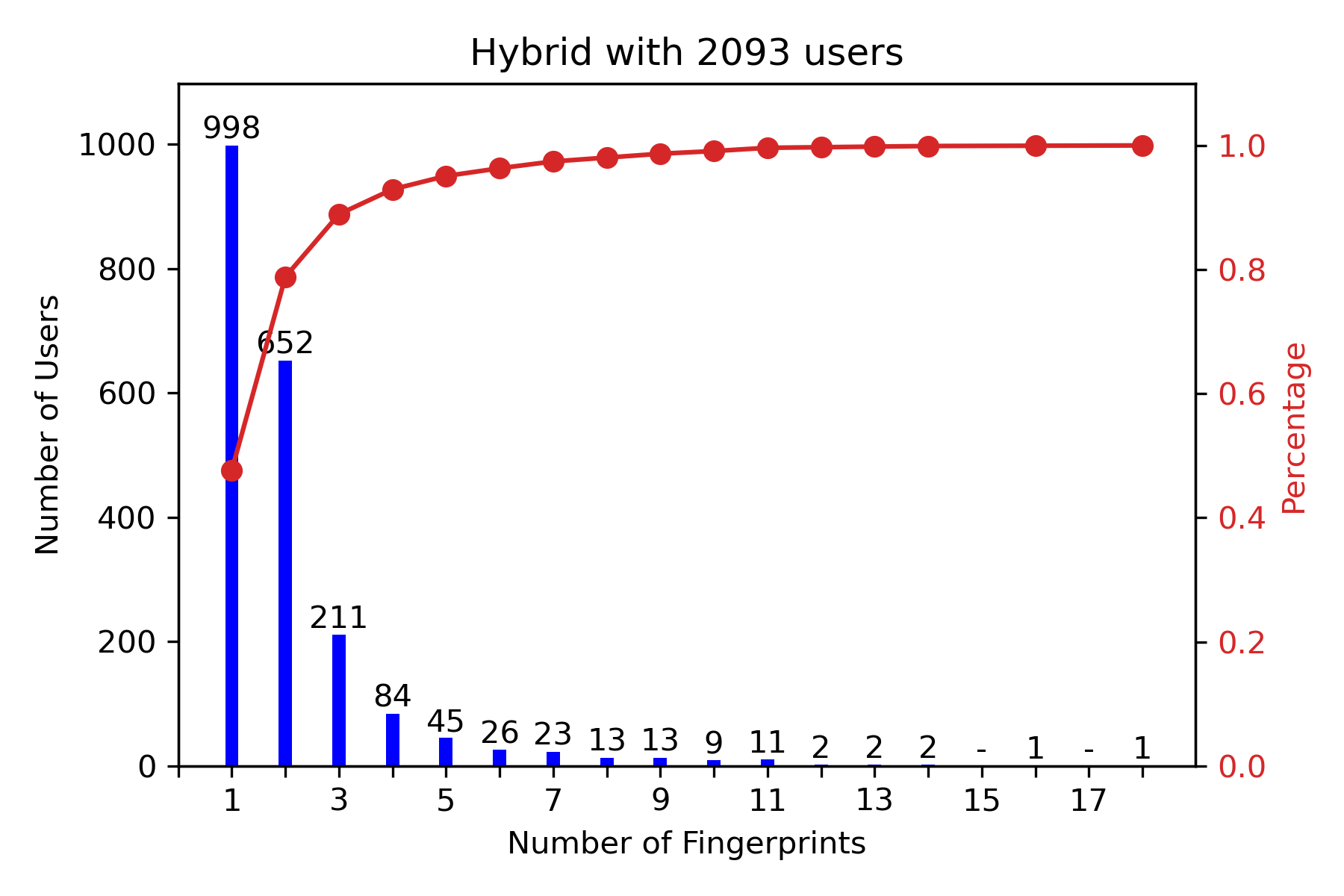}
            \caption{Hybrid (FFT + DC)}
        \end{subfigure}
        \begin{subfigure}[t]{0.49\textwidth}
            \includegraphics[width=\textwidth]{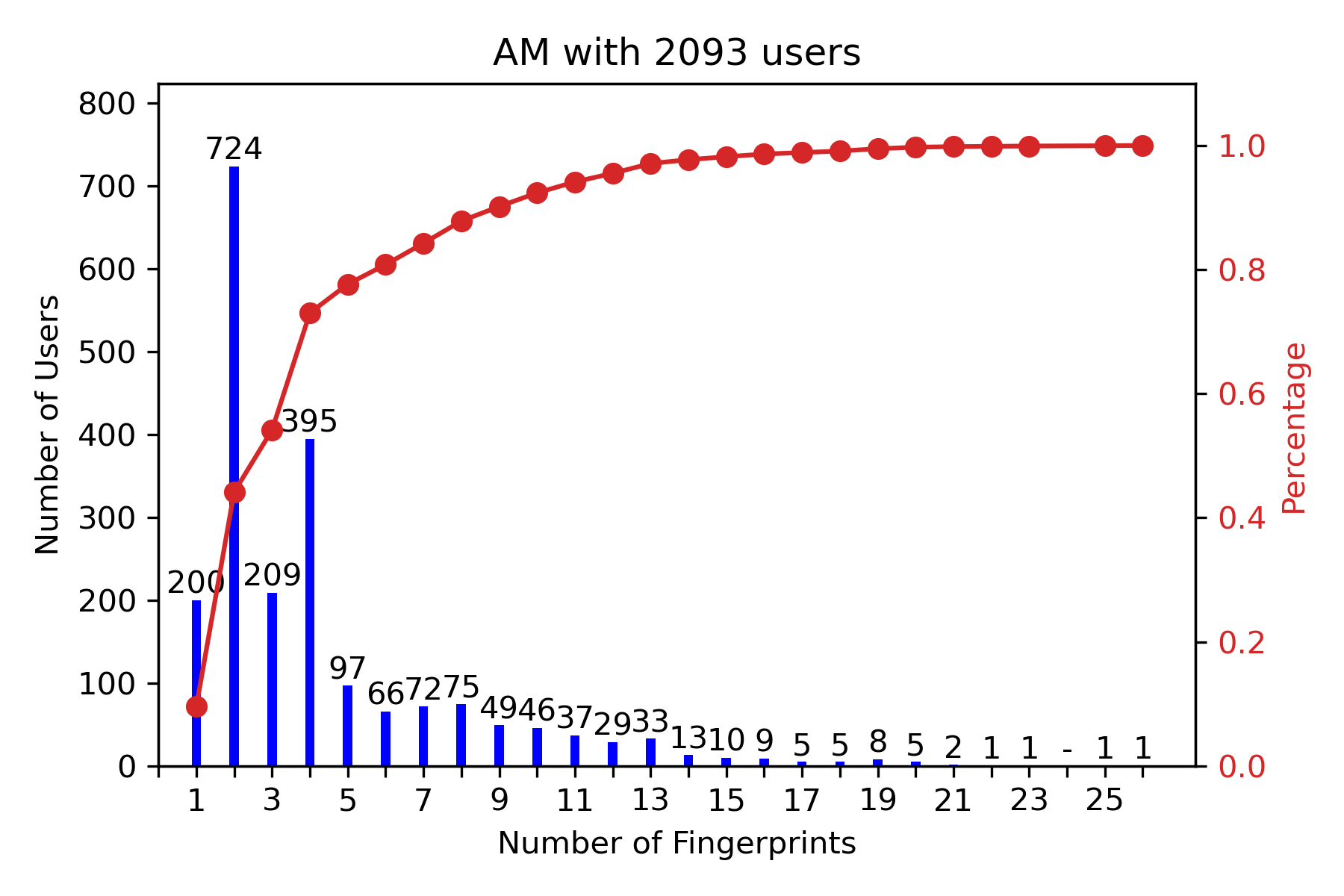}
            \caption{Amplitude Modulation (AM)}
        \end{subfigure}
        \caption{CDFs and Bar plots showing the distribution of number of distinct fingerprints.}
        \label{fig:fp_num_dist}
 \end{figure*}
 


At the same time, the other columns in Table~\ref{table:fp_stats} which show the minimum and mean number of fingerprints obtained from a user's browser reveal that there are users who only left one fingerprint among all 30 iterations. It is to be noted that the ``Min.'' value is 1 for all rows in the table. Furthermore, it is to be noted that the ``Max.'' value for any row in the column is only 26 and not 30 even though the number of iterations of fingerprinting is 30. This shows that there are some fingerprints that are repeating for every vector across every user. Figure~\ref{fig:fp_num_dist} shows the distribution of fingerprint numbers for some of the vectors using both CDFs and bar plots\footnote{The graphs for the remaining four FFT-based vectors are presented in Appendix~\ref{sec:app_stability} due to space limitations.}. The graphs clearly show that the number of distinct fingerprints for most users is simply one or two thus indicating high degree of stability for most users. Furthermore, even with the AM vector which had as many as 26 fingerprints for one particular user, we can notice with the help of the CDF that more than 90\% of users have at most 8 number of distinct fingerprints in the 30 iterations. All of this shows that there is a degree of stability in all of these vectors. Inspired by this, in the next subsection, we devise a simple graph-based fingerprint collation algorithm to combine all the various fingerprints in the 30 iterations into a single fingerprint. We then measure the effectiveness of this approach by measuring the stability of these fingerprinting vectors using that algorithm.

\subsection{Fingerprint Collation via Graphs}
\label{ssec:graph_approach}

\begin{figure}
    \centering
    \includegraphics[width=\columnwidth]{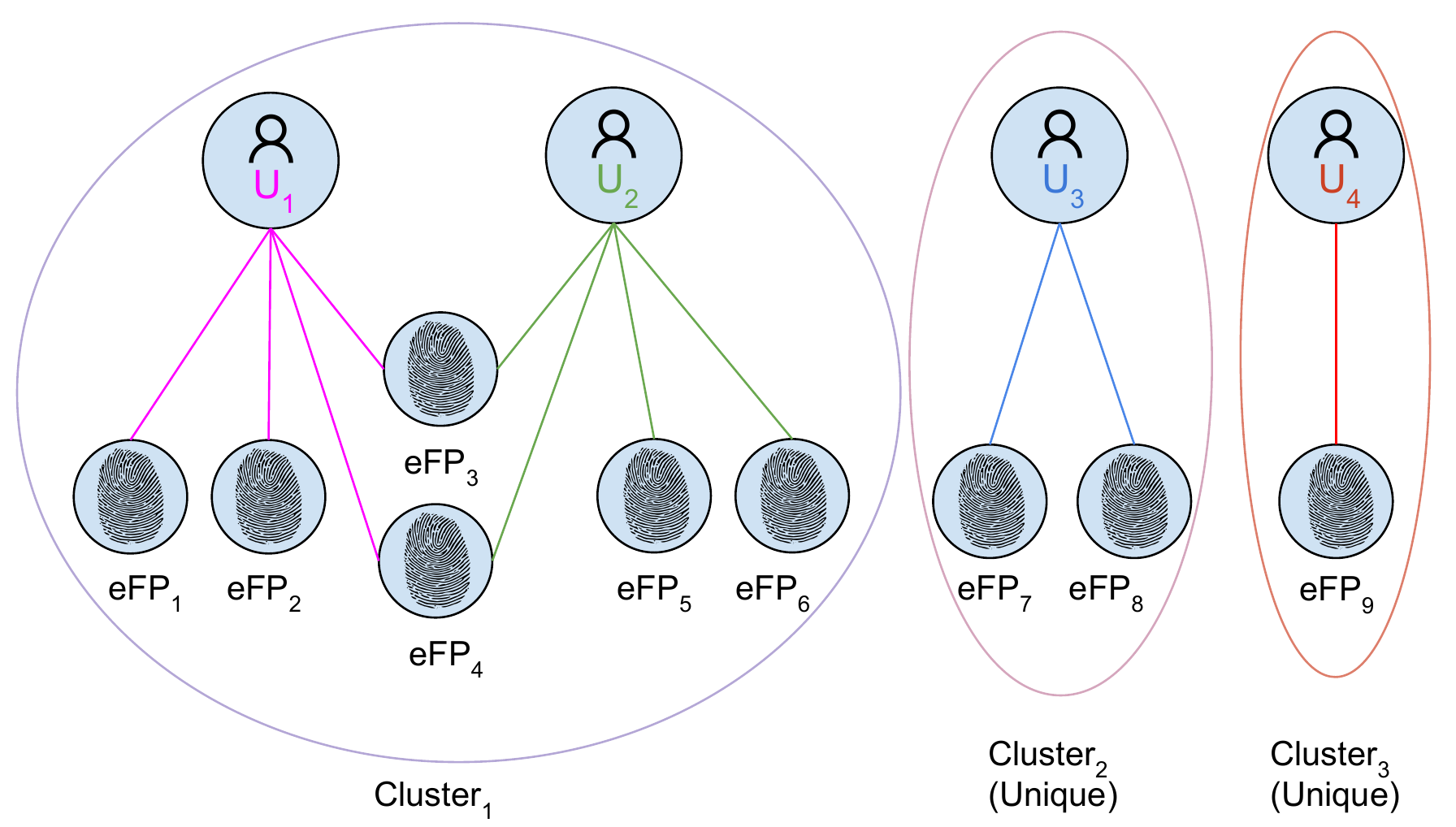}
    \caption{Our graphical approach for collating multiple fingeprints into a single fingerprint}
    \label{fig:graphical_approach}
\end{figure}
In order to aggregate the multiple audio fingerprints that some users have into a single fingerprint, we pursue a simple graph-based approach. For every fingerprinting vector, we build a separate undirected graph in which every user and every elementary fingerprint is represented by a node. For example, Figure~\ref{fig:graphical_approach} represents a hypothetical graph for a particular vector after collecting 9 elementary fingerprints ($eFP_1$ to $eFP_9$) from across 4 users ($U_1$ to $U_4$). In this graph, all fingerprint nodes are connected to all the user nodes that they were associated with during the fingerprint collection process. In order to collate the fingerprints, we simply consider each \emph{connected component} in the graph to be representation of each \textbf{collated fingerprint}. Thus, the number of connected components is the number of distinct collated fingerprints and each user in a particular component can be considered to have the same fingerprint. In our example, we thus end up with 3 distinct fingerprints for the 4 users with users $U_1$ and $U_2$ having the same fingerprint while users $U_3$ and $U_4$ having a unique fingerprint that does not collide with any other user's. Thus each connected component can also be considered to be a cluster of users (\textbf{user cluster}) with colliding fingerprints.

The above graph approach will also work seamlessly for vectors that do not exhibit any apparent ``fickleness''. For example, consider the Dynamics Compressor (DC) vector which resulted in the same fingerprint in each iteration for every user. Each connected component in the graph for such a vector will only have one elementary fingerprint unlike in Figure~\ref{fig:graphical_approach}. Thus, in this case, the proposed approach will function in the same way as the traditional approach of clustering users based on the exact match of their single fingerprint obtained in their visit.

It is to be noted that with our proposed method, as we obtain fingerprints of more users, new collisions can pop up between users who were previously considered to be having distinct fingerprints. For example, consider a new user $U_5$ who has elementary fingerprints, $eFP_8$ and $eFP_9$. This merges existing second and third user clusters into one large cluster that make all three users $U_3$, $U_4$, $U_5$ to be considered to have the same colliding fingerprint. This means that the fingerprinting graph has to be adjusted in a dynamic fashion by the fingerprinter. For this, fingerprinters can rely on prior works such as~\cite{HolmLT01} that proposed fully online graph algorithms for dynamic connectivity queries. The algorithm proposed in \cite{HolmLT01} has an amortized operation cost of $\mathcal{O}(\log^2{}n)$ for graph updates and $\mathcal{O}(\log{}n/\log{}\log{}n)$ for connectivity queries where $n$ is the number of vertices in the graph. Let us assume that a particular fingeprinter has $u$ users fingerprinted with a particular vector where the number of iterations for each user is $k$ (note that $u=2093$ and $k=30$ in our study). In the worst case, even if every fingerprint in every iteration for every user is distinct, the maximum number of nodes in vertices will be $(k+1)u$ as there will be $u$ users and $ku$ fingerprints. Thus, the graph update operation cost for a fingerprinter is only $\mathcal{O}(\log^2{}u)$ while the query operations cost even less. Thus, we can see that this approach scales well to even billions of users. 


\subsection{Stability Analysis}
\label{ssec:stability_analysis}

We have proposed a fingerprint collation approach in order to aggregate multiple fingerprints that were seen for all FFT-based vectors. However, the question of whether this approach results in \emph{stable} fingerprints still remains. It is clear that due to the ``fickle'' nature of FFT vectors it is necessary for the fingerprinting code to be run more than one time (defined as $k$ here). But, it is unclear how much the ideal value of $k$ should be for the various fingerprinting vectors we consider. We attempt to answer both these questions using two measurement approaches (clustering agreement and fingerprint matching scores) which we will describe below.

\subsubsection*{Clustering Agreement Scores} For this, we first break down the fingerprint iterations in our dataset of size $k$ ($=30$) into multiple equal-sized subsets of size $s$. Then, for each vector and a particular value of $s$ ($<k$), we can obtain a clustering of users using the proposed fingerprint collation algorithm. For example, consider the value of $s=10$ which implies that we break down the elementary fingerprints obtained during the 30 iterations into 3 disparate subsets each of size 10. Using only the data from first subset, we obtain a different clustering of users for each audio fingerprinting vector $v$. We can then do the same for the other subsets resulting in a total of $\floor*{\frac{k}{s}}$ clusterings for each vector. We can then use a cluster agreement measuring algorithm to compare how much clusterings from each of the $\floor*{\frac{k}{s}}$ different subsets agree with one another. For measuring cluster agreement, we use the Adjusted Mutual Information (AMI) metric which is an information theoretic measure for clustering comparison~\cite{NguyenEB09}. We chose AMI as it was shown by researchers to be a suitable algorithm for comparing clusters of imbalanced sizes (with small-sized clusters)~\cite{RomanoVBV16} which is typically the case with browser fingerprints~\cite{LaperdrixRB16}. The AMI scores vary between 0 and 1 with 1 indicating exact matching of two user clusterings. 


\begin{figure}[h]
    \centering
    \includegraphics[width=0.75\columnwidth]{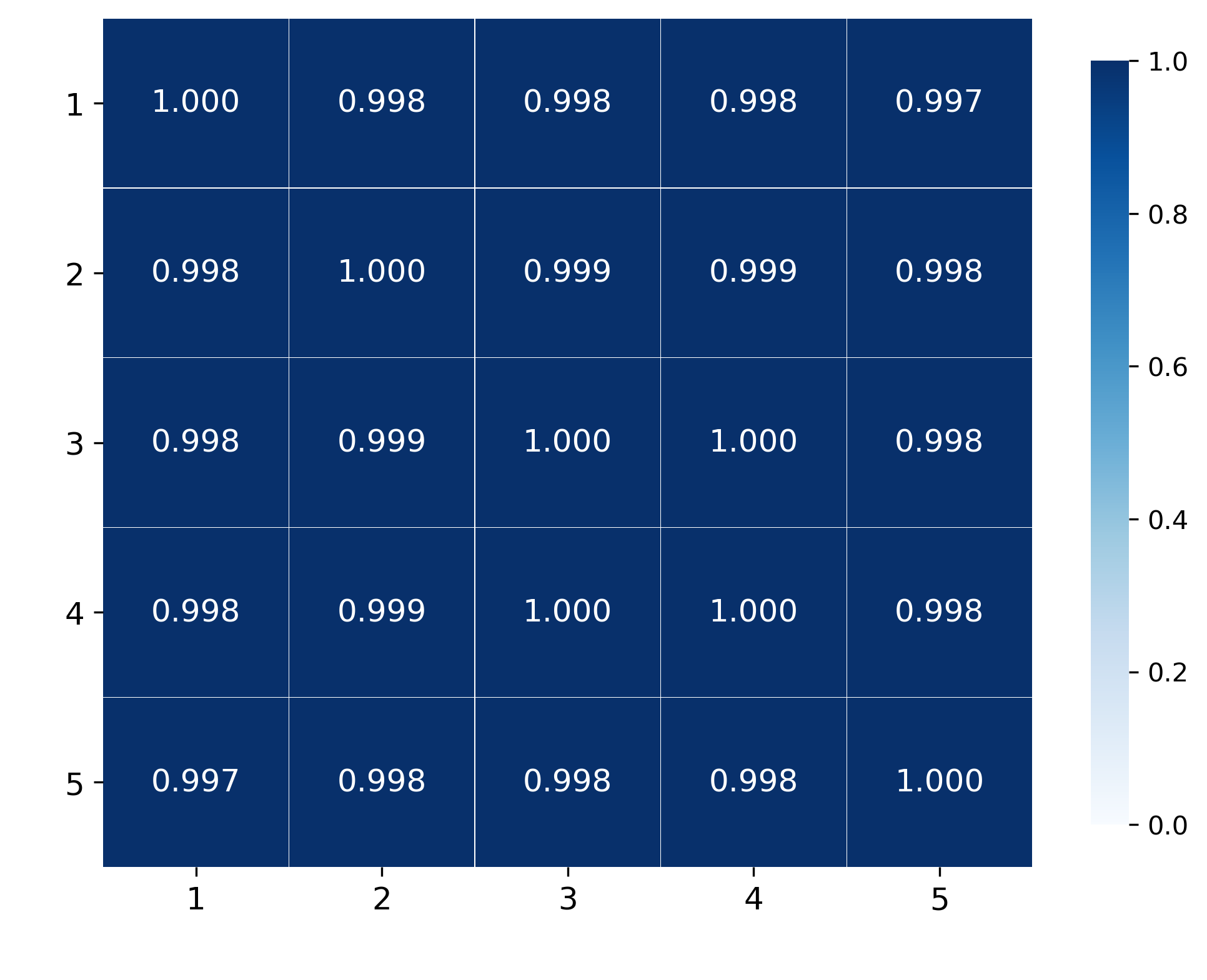}
    \caption{Heatmap depicting cluster agreement scores of FM vector for size $s=6$}
    \label{fig:stability_heatmaps}
\end{figure}

\begin{figure}[h]
    \centering
    \includegraphics[width=1\columnwidth]{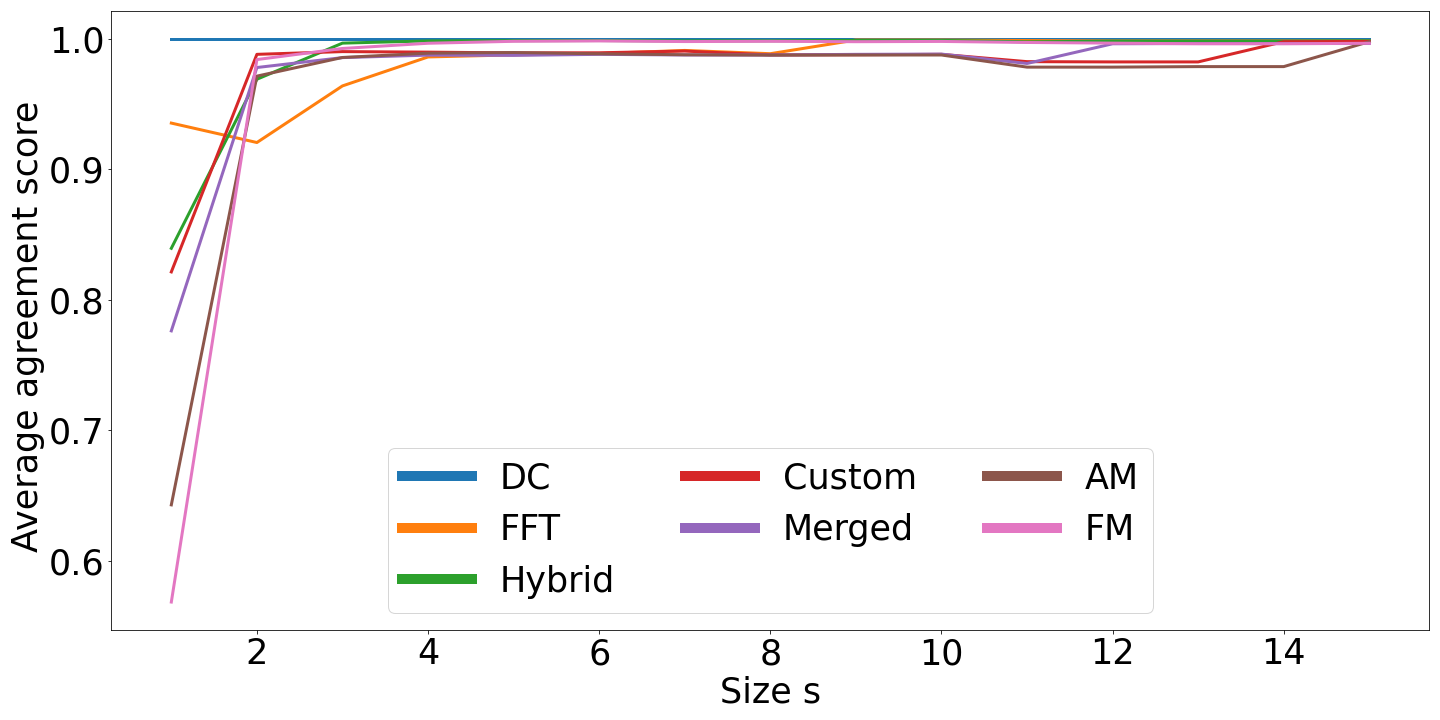}
    \caption{Average cluster agreement scores for different values of $s$ ($[1,15]$) and different vectors.}
    \label{fig:avg_ami_scores}
\end{figure}

A heatmap showing the AMI agreement scores between the 5 different clusterings for size $s=6$ for the Frequency Modulation vector is shown in Figure~\ref{fig:stability_heatmaps}. As shown in the figure, the agreement scores are very high with all scores being more than 0.997 indicating that clusterings obtained from different iterations highly agree with one another. We have repeated these measurements for other values of $s$ as well as other fingerprinting vectors and present the average cluster agreement scores in Figure~\ref{fig:avg_ami_scores}. Note that when $s$ is not a factor of $k$ ($=30$), we simply consider only the first $\floor*{\frac{k}{s}}s$ iterations which are part of the first $s$ subsets and ignore the last few iterations. For $s=4$, the minimum average value of the score is 0.986 (for FFT vector) whereas for $s=15$, this value is 0.997 (for Merged Signals vector). The results clearly show that even for low values of $s$ (as long as it is at least two), the audio fingerprints using  our proposed graph-based collation algorithm result in user clusterings that are highly similar to one another for a given vector across repeated attempts. 

\subsubsection*{Fingerprint Match Scores} The high AMI scores showed that similar clusters are produced across different iterations for each vector. However, it is also vital for a fingerprinter to pinpoint a given visitor to exactly the same user's connected component (or cluster) generated in a prior visit. This is what allows fingerprinting to be consistent across multiple visits of a given user. In order to measure this, we follow a simple procedure outlined here. As before, we divided the fingerprint iterations into subsets of size $s$ ($=3,10,15$). For each value of $s$ and each vector, we consider the first subset as a ``training set'' and use its fingerprints to build a \textbf{training graph} as in Figure~\ref{fig:graphical_approach}. We then consider the elementary fingerprints from each of the remaining subsets iteratively for each user and measure how many of the users can be mapped to the same cluster that they belong to as per the graph built from the first subset. This can be explained with a representative example. Assume that Figure~\ref{fig:graphical_approach} represents a graph that was built for $s=15$ for 4 users. This leaves another 15 iterations of elementary fingerprints (the second subset) for each of the 4 users. Now, consider the user $U_3$'s 15 fingerprints from this second subset. There can be following potential example cases regarding $U_3$:
\begin{enumerate}
\item If $U_3$'s fingerprints (from the second subset) happen to be only $eFP_7$, $eFP_{10}$, then the user will be pointed to the right cluster (Cluster 2) and hence can be considered as a ``positive outcome''.
\item If $U_3$'s remaining fingerprints are $eFP_10$ alone then the user will not be pointed to any cluster and will instead result in creation of a new cluster due to lack of prior connections. Hence, this can be considered as a ``negative outcome''.
\item Another case is  If $U_3$'s fingerprints are $eFP_7$ and $eFP_9$ thus forcing us to merge two previously distinct clusters. Again, this will be considered a ``negative outcome'' as we fail to uniquely identify the one right cluster (Cluster 2) for the user being considered.
\end{enumerate}

Using the above simple approach, we computed the fraction of remaining ``user subsets'' which we were able to positively point to the right cluster in the training graph. These results are presented in Table~\ref{table:fp_match_scores} and show that even for very small set sizes we are able to accurately point the vast majority of users uniquely to their ``original cluster'' based on their current fingerprints. This establishes the ability of web audio API-based code to produce persistent fingerprints despite their apparent fickleness that was originally seen. 

\begin{table}[ht]
\small
\centering
    \begin{tabular}{lccc}
    \toprule
    {\bf Fingerprinting Vectors} & {\bf $s=15$}& {\bf $s=10$}& {\bf $s=3$} \\
    \midrule
    DC & 1.0 & 1.0 & 1.0 \\ 
    \midrule
    FFT & 1.0 & 1.0 & 0.9942 \\ 
    \midrule
    Hybrid (DC + FFT) & 1.0 & 1.0 & 0.9952 \\ 
    \midrule
    Custom Signal & 0.999 & 0.9988 & 0.9969 \\ 
    \midrule
    Merged Signals & 1.0 & 0.9998 & 0.9953 \\ 
    \midrule
    AM & 0.999 & 0.9983 & 0.991 \\ 
    \midrule
    FM & 0.9981 & 0.9978 & 0.9899 \\ 
    \midrule
    \end{tabular}
    \caption{Fingerprint match scores.}
    \label{table:fp_match_scores}
\end{table}

\subsection{Timing Analysis}
\label{ssec:timing}
Besides stability, we also measured the time it takes for fingerprinting to happen. The time taken to run one entire iteration of the fingerprinting code stack for all the vectors is about 1 second with an average time of only about 0.14 seconds per vector. These low run times show that a fingerprinter can potentially run multiple iterations of any of the effective audio fingerprinting vectors without worrying about time constraints.

\section{Diversity Analysis}
\label{sec:diversity}



Entropy measures are commonly used to measure the diversity and there by, the ``fingerprinting power'' of browser fingerprints~\cite{LaperdrixRB16,GomezLB18,LaperdrixBBA20}. We followed the same approach for our study and computed the Shannon bit entropy as well as normalized entropy for all the web audio fingerprinting vectors that we studied. We describe the computation here for clarity. Assume that there exist $n$ distinct fingerprints, with $u_i$ (where $i \in [1,n]$) denoting number of users in the study that have the $i$th fingerprint and $U$ denoting total number of users. We compute bit entropy $\boldsymbol{e}$ for a given fingerprinting vector as below:
\[
      e = - \sum_{i=1}^{i=n} \frac{u_i}{U} \log_{2}  \frac{u_i}{U}
\]

Then, the normalized entropy ($\boldsymbol{e_{norm}}$) is obtained by dividing the bit entropy by the maximum possible entropy i.e. $\frac{e}{\log_{2}{U}}$ in order to bring it down to a range of 0 to 1. Note that 1 indicates maximum possible entropy and unique fingerprintability of every user. This normalized measure enables comparison between fingerprint fingerprinting measures of various studies even if the number of users in the study is different~\cite{LaperdrixRB16,GomezLB18}. 
 
We will first discuss the diversity of Web Audio configuration information gleaned from the users browsers. As in~\cite{LaperdrixBM17}, we found this information to be less useful for fingerprinting. For example, we have only found 5 distinct values of default \texttt{sampleRate} with 48 KHz being the most common value (76.8\%) followed by 44.1 KHz (22.7\%). The rest of the 3 values were only found among 9 users in our study. This echoes the observations made by browser developers in the Privacy Interest Group (PING) as well when discussing the documentation for Web Audio API~\cite{ping}. Similar is the case with the \texttt{maxChannelCount} parameter that only had 6 distinct values. However, interestingly, we noticed that an experimental property named \texttt{AudioContext.baseLatency} which documents the incurred processing latency proved to be the most diverse configuration property with 29 distinct values and 10 unique values for the 2093 users. Although 60\% of the users had a base latency of 0.01 seconds, other values ranging from 0 to 0.16 seconds were also seen in the data. Overall, this property had an entropy of 1.5 bits with a normalized entropy of about 0.198. While~\texttt{baseLatency} has been recently considered by PING as a candidate for fingerprinting~\cite{jasonnovak_2018}, our results here help quantify the privacy threat.  

The diversity of the 7 more advanced dynamic fingerprint vectors based on utilizing Web Audio APIs  is presented in Table~\ref{table:diversityanalysistable}. In order to allow for comparison, in Table~\ref{table:diversityothers} we also present the entropy values of other fingerprinting vectors which were shown to be effective in prior works. Table~\ref{table:diversityanalysistable} shows that FFT-based audio vectors are more effective at fingerprinting than pure Dynamics Compressor vector with a normalized Shannon entropy of more than 0.23. Most of these FFT-based vectors result in 80-85 distinct fingerprints for the 2093 users with about 40 of them being unique (i.e. only associated with one user in the dataset). The table shows that all the diversity values of the FFT-based vectors are very close to one another thus indicating that the discriminatory cause behind all these vectors is potentially the FFT operation alone. 

The final row of Table~\ref{table:diversityanalysistable} considers a combination of all the individual audio fingerprints. In order to compute the diversity of the combination of multiple fingerprinting vectors, the following simple logic is used. Assume, that a user $U_i$ has multiple fingerprints associated with different vectors such as $f_i$, $g_i$, $h_i$ etc. Then, in order to find the diversity of a combination vector of all these individual vectors, we simply compute the diversity of tuples: $(f_i, g_i, h_i, ...)$ across all values of $i$. By definition, the diversity of a combination vector will at least be as much as the diversity of the most diverse component vector. We can see in Table~\ref{table:diversityanalysistable} that the entropy of combinations of all audio vectors is again close to that of the FFT-based vectors thus providing further proof for alignment of all FFT-based vectors. We will discuss this more in Section~\ref{sec:discussion}. 

Comparing Tables~\ref{table:diversityanalysistable} and \ref{table:diversityothers} shows that the diversity of audio fingerprints is much less than that of other effective fingerprinting vectors such as Canvas, Fonts and \texttt{User-Agent} header based fingerprints. This difference can also be seen in terms of number of distinct and unique fingerprints. 

\begin{table}[ht]
\small
\centering
    \begin{tabular}{lccccc}
    \toprule
    {\bf Vectors} & {\bf Distinct} & {\bf Unique} & {\bf Entropy} & $\boldsymbol{e_{norm}}$\\
   \midrule
    DC & 59 & 34 & 1.935 & 0.175 \\ 
    \midrule
    FFT & 73 & 42 & 2.593 & 0.235 \\ 
    \midrule
    Hybrid & 84 & 42 & 2.692 & 0.244 \\ 
    \midrule
    Custom Signal & 72 & 41 & 2.582  & 0.234  \\ 
    \midrule
    Merged Signals&  87 & 45 & 2.767 & 0.251 \\ 
    \midrule
    AM & 82 & 45 & 2.69 & 0.244 \\ 
    \midrule
    FM & 82 & 43 & 2.717 & 0.246 \\ 
     \midrule
	\midrule
      Combined & 95 & 49 & 2.803 & 0.254 \\ 
     \bottomrule

    \end{tabular}
    \caption{Diversity of audio fingerprints (2093 users)}
    \label{table:diversityanalysistable}
\end{table}

\begin{table}[ht]
\small
\centering
    \begin{tabular}{lccccc}
    \toprule
    {\bf Vectors} & {\bf Distinct} & {\bf Unique} & {\bf Entropy} &$\boldsymbol{e_{norm}}$\\
   \midrule
   Canvas & 352 & 224 & 6.109 & 0.554 \\ 
    \midrule
    Fonts & 690 & 555 & 7.146 & 0.648 \\ 
    \midrule
    \texttt{User-Agent} & 427 & 284 & 6.466 & 0.586 \\ 
     \bottomrule
    \end{tabular}
    \caption{Diversity of other vectors (2093 users)}
     \label{table:diversityothers}
\end{table}

\subsubsection*{Comparison with User-Agent fingerprints} 
\begin{figure*}[h]
    \centering
    \includegraphics[scale=0.3]{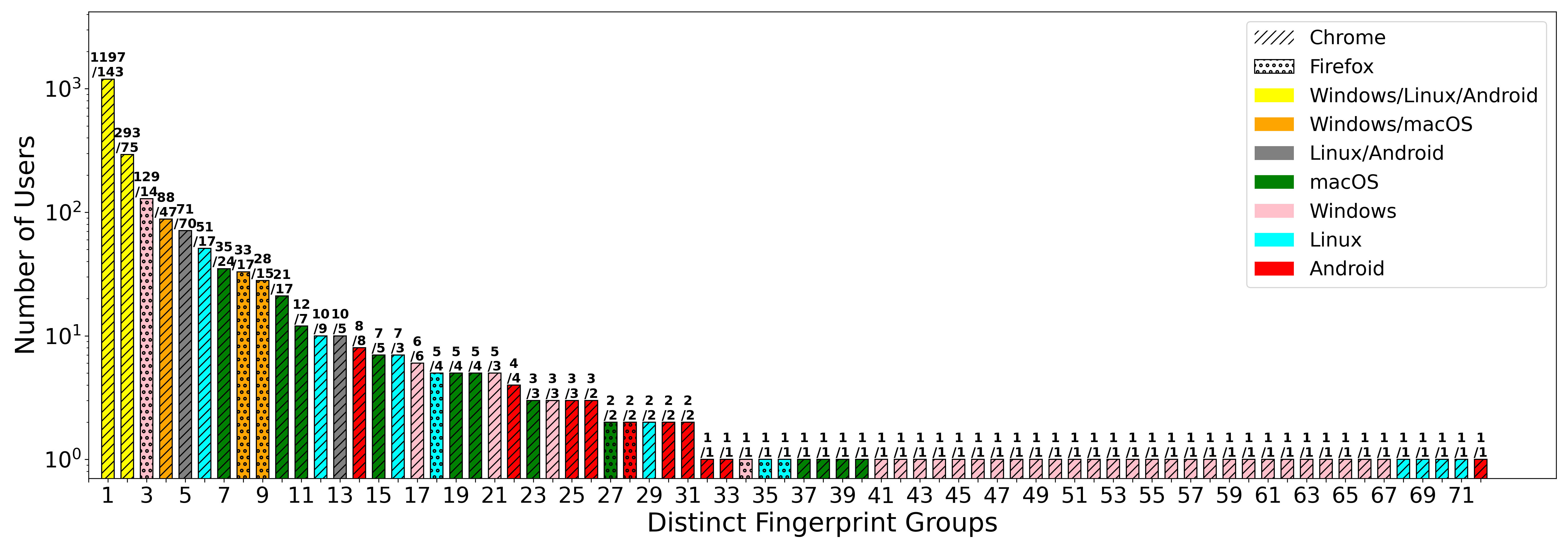}
    \caption{User counts and browser/OS types for Custom Signal clusters}
    \label{fig:ua_breakdown_custom}
\end{figure*}

The~\texttt{User-Agent} (UA) header is an indicator of the web browser, its version number as well as the OS being used to visit a web server. As browser fingerprints typically change across different UAs, it would be insightful to compare audio fingerprints with UAs. In order to do this, we first considered the user clusters produced by each of the fingerprint vectors and evaluated the homogeneity of these clusters in terms of the UA strings. Figure~\ref{fig:ua_breakdown_custom} depicts the distribution of users as well as the distinct browser/OS counts (as inferred from the UAs) across various user clusters produced by the Custom Signal vector\footnote{The graphs for the remaining 6 vectors are included in Appendix~\ref{sec:fp_breakdown_by_ua}}. Two different fill patterns indicate browser families while different colors indicate different browser/OS combinations for the users in each particular user cluster. The numbers at the top of each bar indicate the number of users and the number of distinct UA strings for users in each cluster.

Multiple observations can be made from this distribution graph. Firstly, the distribution of users shows that a minor fraction of fingerprints account for a major portion of the users with top 3 popular fingerprints accounting for more than 75\% of the users. This is very common with browser fingerprinting mechanisms as was documented in prior works~\cite{LaperdrixRB16}. It should be noted that neither browser family (Firefox nor Chrome) is pre-disposed towards either unique fingerprints or popular fingerprints. For example, the graph shows that about 51 users are associated with 1 or 2 size clusters for Custom Signal vector. Among these 51, we can see that about 7 users are using Firefox where as the rest 44 are Chrome-based which also resembles the base ratio of Firefox and Chrome users. More interestingly, one can see a 100\% homogeneity in the clusters in terms of the browsers as there exists no user cluster that has both Chrome and Firefox fingerprints. On the other hand, for both Firefox as well as Chrome, there do exist user clusters which are associated with more than one OS (heterogeneous) as indicated by the yellow, orange and gray colored bars in Figure~\ref{fig:ua_breakdown_custom}. This indicates that~\emph{the browser code is more important than the OS as a differentiating factor for web audio fingerprints}.

Encouraged by the observed browser homogeneity in the user clusters, we wanted to verify a related assertion made by the World Wide Web Consoritum's (W3C) organization's standards document stating that Web Audio fingerprinting ``merely allows deduction of information already readily available by easier means (User Agent string)''~\cite{wwwc_2021}. For this, we first considered only the 143 UA strings that were each associated with more than one user in our dataset. These 143 strings were all seen with 1950 users in total in our study. Of these 143 UAs, we noted that as many as 90 of them were spanning multiple FFT-based fingerprint clusters\footnote{This number of 90 is about same for all the 6 FFT-based vectors.}. Together, these accounted for about 1610 of the 1950 users. Further, several of these UAs were associated with more than 2 fingerprint clusters. For example, 7 UAs were each associated with at least 5 different Merged Signal fingerprints with one particular Chrome/Windows UA being associated with as many as 10 different fingerprints. However, we did not notice any explicit differences between browser families in this behavior with both Firefox and Chrome UAs both getting frequently associated with more than one audio fingerprint. This clearly shows that~\emph{unlike what was mentioned in W3C's documentation, there are a significant number of cases where audio fingerprinting reveals more information about users than \texttt{User-Agent} fingerprinting alone}.


\subsubsection*{Additive Value of Audio Fingerprints}

\begin{table}[ht]
\small
\centering
    \begin{tabular}{lccccc}
    \toprule
    {\bf Vectors} & {\bf Distinct} & {\bf Unique} & {\bf Entropy} &$\boldsymbol{e_{norm}}$\\
   \midrule
   Canvas & 352 & 224 & 6.109 & 0.554 \\ 
   \midrule
   Canvas + Audio & 492 & 318 & 6.699 & 0.607 \\ 
   \midrule
   \midrule
    Canvas + Font & 1106 & 916 & 9.086 & 0.824 \\ 
    \midrule
    Canvas + Font + Audio & 1210 & 1010 & 9.351 & 0.848 \\ 
    \midrule
   \midrule
    Canvas + Font + UA  & 1640 & 1436 & 10.422 & 0.945 \\ 
    \midrule
     Canvas + Font  & 1680 & 1493 & 10.479 & 0.95 \\ 
    + UA + Audio &&&& \\
    \bottomrule
    \end{tabular}
    \caption{Assessing additive value of audio vectors}
    \label{table:additive_value}
\end{table}

The above showed that web audio fingerprints have more fingerprinting value beyond simply recording the~\texttt{User-Agent} header. It would be useful to quantify this additive value that audio fingerprinting can potentially add to existing powerful fingerprinting schemes. For this, we first consider Canvas fingerprinting as it was shown to be one of the most discriminative fingerprinting techniques previously~\cite{LaperdrixRB16}. We measured the entropy of a ``pure'' Canvas API-based fingerprinting technique as well as ``Canvas + Audio'' fingerprint where Audio fingerprint includes an aggregations of all 7 web audio fingerprinting techniques as described previously and shown in the final row of Table~\ref{table:diversityanalysistable}. Table~\ref{table:additive_value} shows that~\emph{audio fingerprinting helps cause a 9.6\% increase in the normalized entropy of Canvas fingerprinting techniques}. The table also shows that this trend of increase in entropy persists even when considering other powerful fingerprinting techniques such as Font and User-Agent-based fingerprinting although the increase in entropy decreases as we keep considering more fingerprinting vectors.\footnote{It is likely that the normalized entropy values for all vectors will decrease as we increase the user study size to the order of millions \cite{LaperdrixBBA20}}. It is to be noted however that Audio fingerprinting (like Canvas fingerprinting) is more difficult to defend against unlike other techniques such as Font and \texttt{User-Agent} fingerprinting. The latter can be tackled by simply changing fonts/\texttt{User-Agent} headers (using a browser extension such as~\cite{ua_switcher}) periodically in a browser. However, combating Canvas and Audio fingerprinting techniques requires more intricate measures such as those taken up by the Brave Browser recently~\cite{brave_audio_fp_defense,LaperdrixBM17} which can have considerable computational as well as compatibility side-effects~\cite{brave_canvas_compat,brave_canvas_compat2} as discussed in Section~\ref{sec:introduction}. We also repeated this analysis for ``UA + Audio'' and saw that it resulted in a $e_{norm}$ value of 0.643,~\emph{a 9.7\% increase from using just UA as a fingerprint thus reaffirming the additive value of audio fingerprinting to UA fingerprinting}.

\textbf{Diversity Results Summary.} Overall, our results show that the privacy threat from standalone web audio fingerprinting is not as serious as it is from some other powerful fingerprinting vectors such as Canvas fingerprinting. At the same time, we showed that audio fingerprinting can act as a significant supplement to existing fingerprinting techniques especially given that it requires intricate measures that carry web compatibility risks to defend against unlike other vectors such as~\texttt{User-Agent} headers. Given this, our study's results will help browsers developers to quantitatively analyze the relative privacy threat posed by audio fingerprinting and make individual design decisions accordingly.


\section{Discussion}
\label{sec:discussion}

\subsubsection*{Participant Pool Size}
Due to financial limitations, we had to restrict the size of our study to 2093 users who were mainly recruited and paid via Amazon's MTurk platform. However, it is important to note that the normalized Shannon entropy measures that we obtained for some well known fingerprinting vectors such as Canvas and~\texttt{User-Agent} are in line with the figures from prior studies that employed even more number of users. For example, the normalized entropy for \texttt{User-Agent} headers in~\cite{LaperdrixRB16} which employed 118,934 users is 0.580 while it is 0.586 in our study. 

Furthermore, we also performed additional analysis to see how our dataset sizes can affect the relative rankings we present. For this, we divided our set of users into 4 disparate equal sized subsets and repeated the entropy analysis for each subset. We noticed that the relative rankings (by $e_{norm}$) of the 9 fingerprinting vectors we covered in Tables~\ref{table:diversityothers} and~\ref{table:additive_value} remained ~\emph{exactly the same across all the small subsets as well as our main dataset}. The results for subsets are given in Appendix~\ref{sec:app_vector_splits}. This further confirms that the analysis we present in our paper remains the same irrespective of the size of the user set that is considered. 


\subsubsection*{Possibility of Other Vectors} In this research, we considered 7 fingerprinting techniques that utilize various APIs offered by the Web Audio standards supported by most modern web browsers. However, one might argue that there might exist some other web audio fingerprinting vector which can potentially be more discriminative than those that we study in this paper. While we concede this is true, we argue that it is very difficult, if not impossible, to exhaustively search for all potential fingerprinting vectors. By varying the choice and order of different API calls to be made as well as number and values of inputs given to the calls, one can potentially generate innumerable audio-based fingerprinting vectors. Furthermore, the same applies to other API-based fingerprinting vectors such as Canvas and WebGL which have been also been studied by focusing on a few specific vectors similarly in prior works~\cite{LaperdrixRB16,GomezLB18}. We therefore took this limitation into consideration and limited ourselves to only 7 vectors which include previously proposed and in-the-wild discovered vectors as well as 4 new vectors that target different web audio APIs to add variety.

\begin{figure}[h]
    \centering
    \includegraphics[width=1\columnwidth]{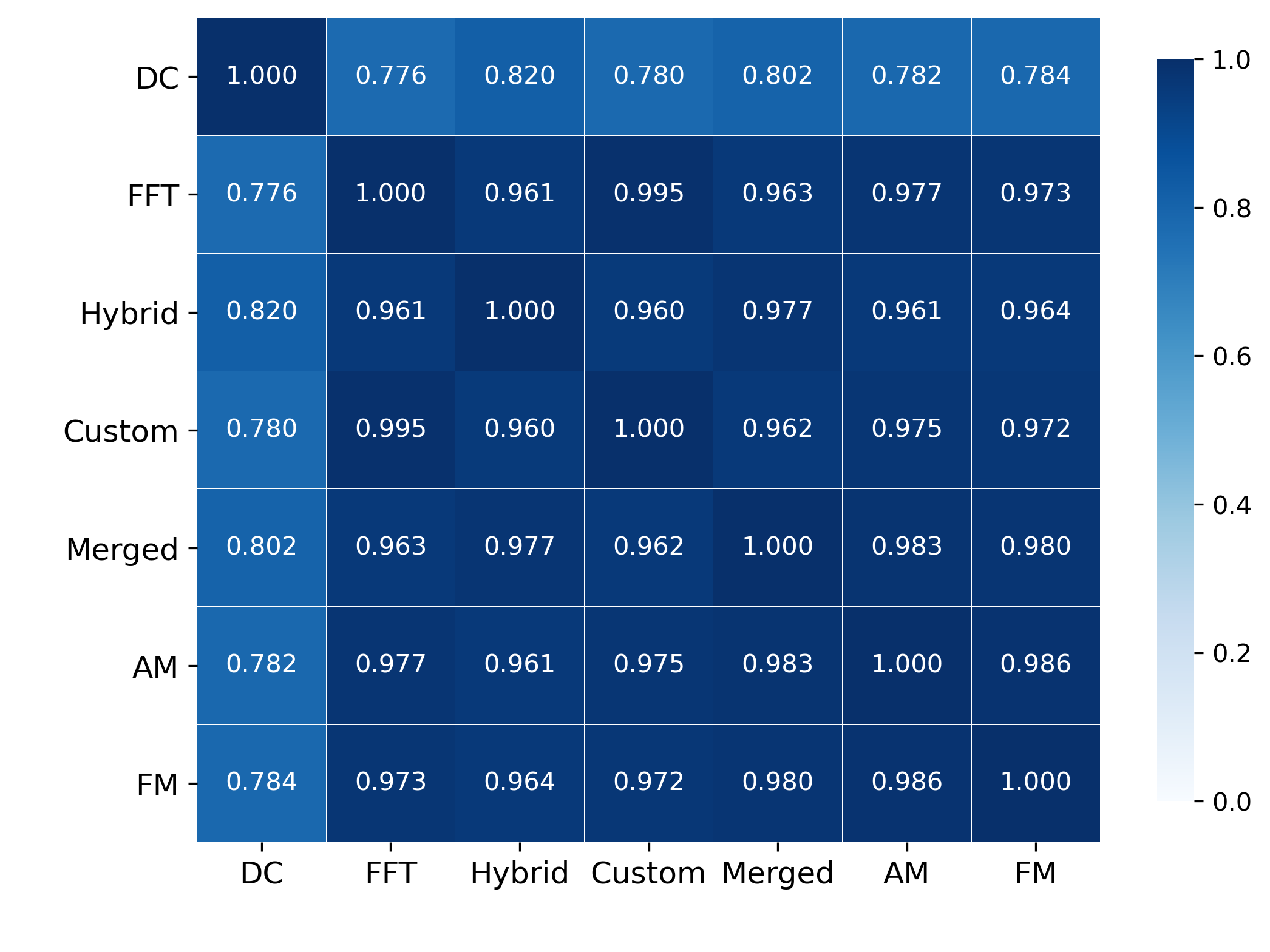}
    \caption{Cluster agreement scores between the different audio fingerprinting vectors.}
    \label{fig:vectors_ami_heatmap}
\end{figure}

Furthermore, our diversity results for all the FFT-based vectors are really close to one another (Table~\ref{table:diversityanalysistable}). We tried to further confirm the alignment of various FFT-based vectors by some additional analysis. For this, we performed a comparison of the user clusters produced by all the 7 fingerprinting vectors (and all $s=30$ iterations) using the Adjusted Mutual Information (AMI) that we described previously. The heatmap describing the results which is presented in Figure~\ref{fig:vectors_ami_heatmap} clearly shows a high agreement between all the FFT-based vectors with AMI scores of at least 0.96. This analysis indicates that the~\emph{FFT operation could potentially be the main discriminative agent in audio fingerprinting code}. Recent discussions by browser developers in W3C's Privacy Interest Group (PING) shed some light on reasons for this as they mention that the floating point Math involved in implementing these APIs could be responsible for the fingerprinting surface of Web Audio APIs~\cite{ping}. We leave further investigation of this to future work.

\subsubsection*{Disclosure}
As discussed in Section~\ref{sec:diversity}, some of our results regarding the diversity of audio fingerprints clearly contradict the web API standards documentation. Prior to publication, we will disclose all our results to the Web Audio Working Group in order to request the documentation's ``Security and Privacy Considerations'' subsection~\cite{wwwc_2021} be updated to accurately delineate the potency of web audio fingerprinting attacks as measured in this study.

\section{Related Work}
\label{sec:related_work}

Browser fingerprinting has received a lot of attention from the research community thus far. Over the years, multiple works have focused on devising fingerprinting techniques~\cite{MoweryS12,nikiforakis13,CaoLW17}. Many works have also focused on measuring and comparing the effectiveness as well as evolution of various browser fingerprints~\cite{LaperdrixRB16,VastelLRR18,GomezLB18}. Further, several studies have also focused on defending against browser fingerprinting attacks~\cite{nikiforakis15,torres15,laperdrix15,LaperdrixBM17,datta19,iqbal20}. However, audio fingerprinting measurements have remained a notable absence in this body of literature. Only \cite{EnglehardtN16} who first discovered audio fingerprinting in the wild and \cite{LaperdrixBM17} have briefly touched upon diversity aspects of audio fingerprinting with a Dynamics-Compressor (DC) vector by conducting user studies. However, to the best of our knowledge, no other work except~\cite{Queiroz} thus far has conducted a dedicated full-fledged analysis of the feasibility and effectiveness of web audio fingerprinting. 


Hence, our closest related work is \cite{Queiroz} by Queiroz et al. In this work, the authors  first manually studied the stability of audio fingerprinting by using  FFT and Dynamics-Compressor (DC) schematics similar to the ones we used in our paper with the help of four personal devices. Unfortunately, based on the apparent ``fickleness'' in the fingerprints exhibited by the FFT vectors, the authors decided to only use pure DC fingerprinting vectors for further evaluation (with 122 devices and 4 separate \texttt{OscillatorNode} signals). However, as we demonstrated with our proposed graph-based approach, FFT-DC vectors can be used as stable fingerprinting vectors with superior diversity results compared to a pure DC vector. Even more importantly,~\cite{Queiroz} does not include measures of the relative importance as well as the additive value of audio fingerprinting when compared to previously known fingerprinting vectors. This is vital as it serves as a measurement to ultimately gauge how effective and useful web audio fingerprinting is to an attacker. 
\section{Conclusion}
\label{sec:conclusion}

In this paper, we conducted the first systematic study of effectiveness of Web Audio-based browser fingerprinting vectors. Firstly, we designed and implemented 4 new audio fingerprinting vectors that made use of Fast Fourier transformations of modulated custom waveforms. We then collected basic web audio configuration information, 7 Web Audio API-based fingerprints as well as multiple other well known browser fingerprints via an elaborate worldwide user study involving 2093 users. After a preliminary analysis of the collected data, we designed a graph-based fingerprint mechanism to collate the multiple audio fingerprints associated with each user. Using this mechanism, we demonstrated that Web Audio APIs can be utilized to yield a stable browser fingerprinting system. After proving the feasibility of audio fingerprinting, we presented detailed diversity measures of audio fingerprints. We showed the relative effectiveness of these fingerprints in comparison to other browser fingerprinting vectors such as Canvas, Font and \texttt{User-Agent}-based fingerprinting to help future browser developers to take informed design decisions regarding privacy protection.

\bibliographystyle{ACM-Reference-Format}
\bibliography{reference}

\clearpage
\appendix
\section{Distribution of Audio Fingerprints across Users}
\label{sec:app_stability}

        \begin{figure}[h]
          \centering
            \includegraphics[width=\columnwidth]{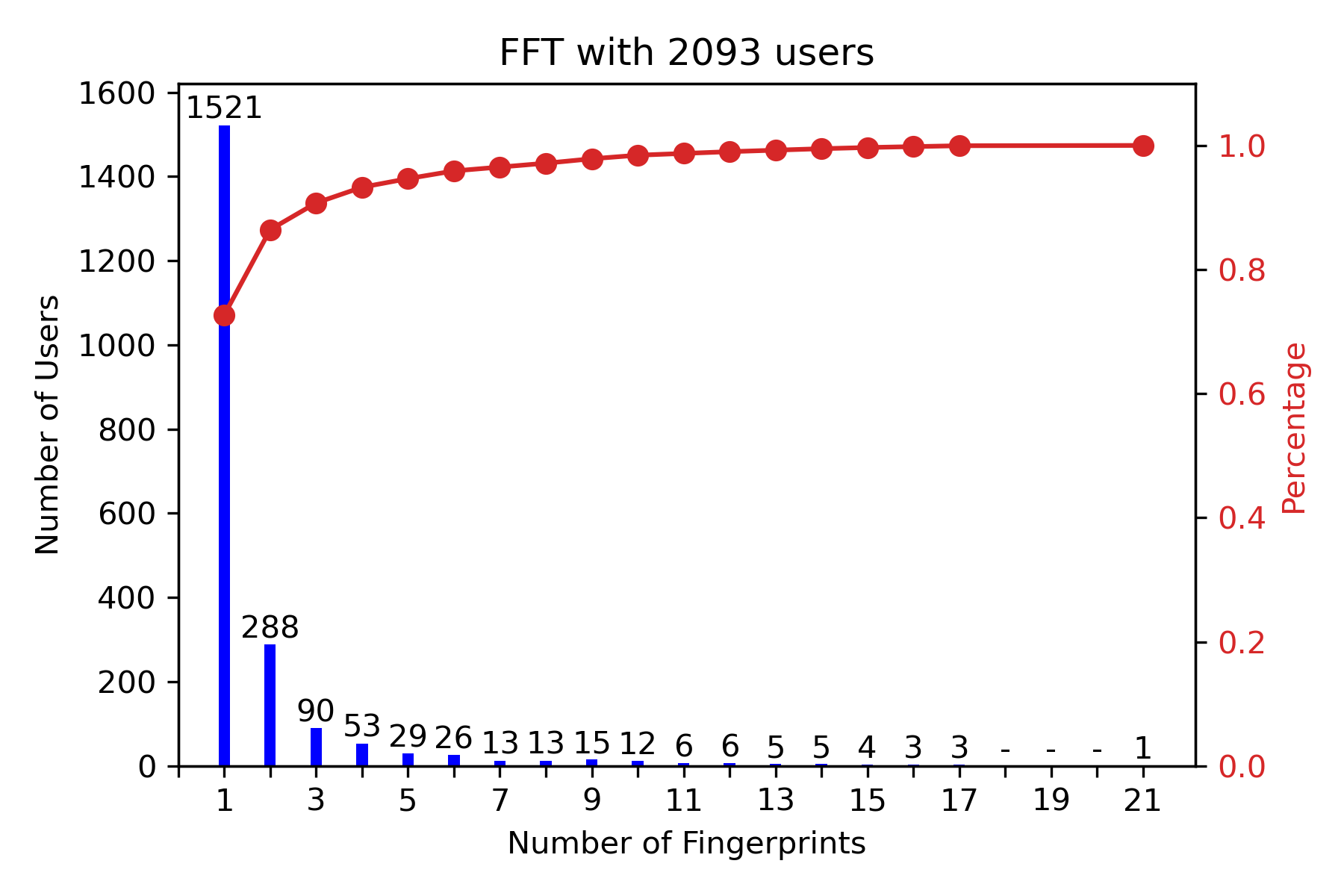}
            \caption{CDFs and Bar plots showing the distribution of number of FFT fingerprints.}
        \end{figure}
        
         \begin{figure}[h]
          \centering
            \includegraphics[width=\columnwidth]{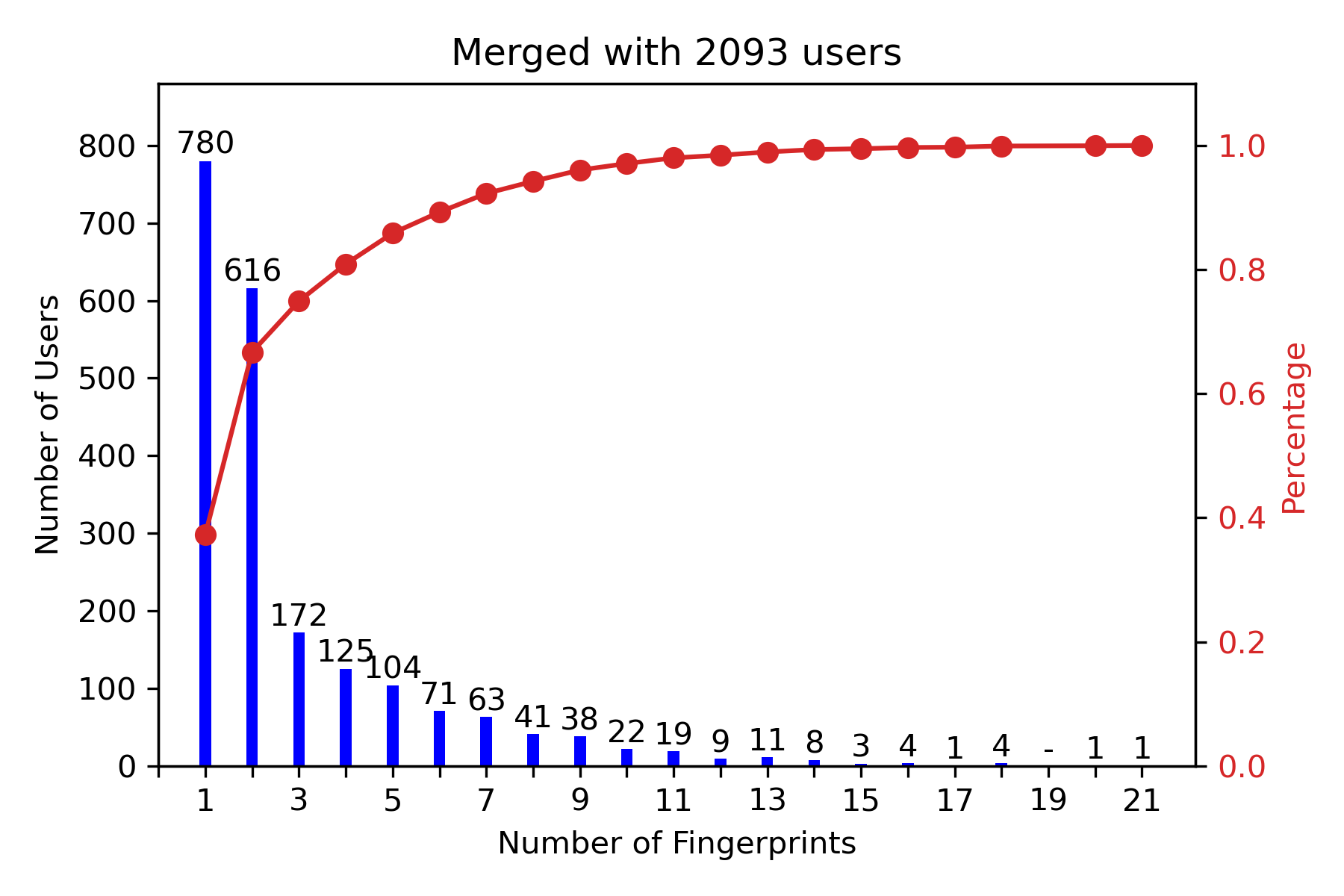}
            \caption{CDFs and Bar plots showing the distribution of number of Merged Singals fingerprints.}
        \end{figure}
        
         \begin{figure}[h]
          \centering
            \includegraphics[width=\columnwidth]{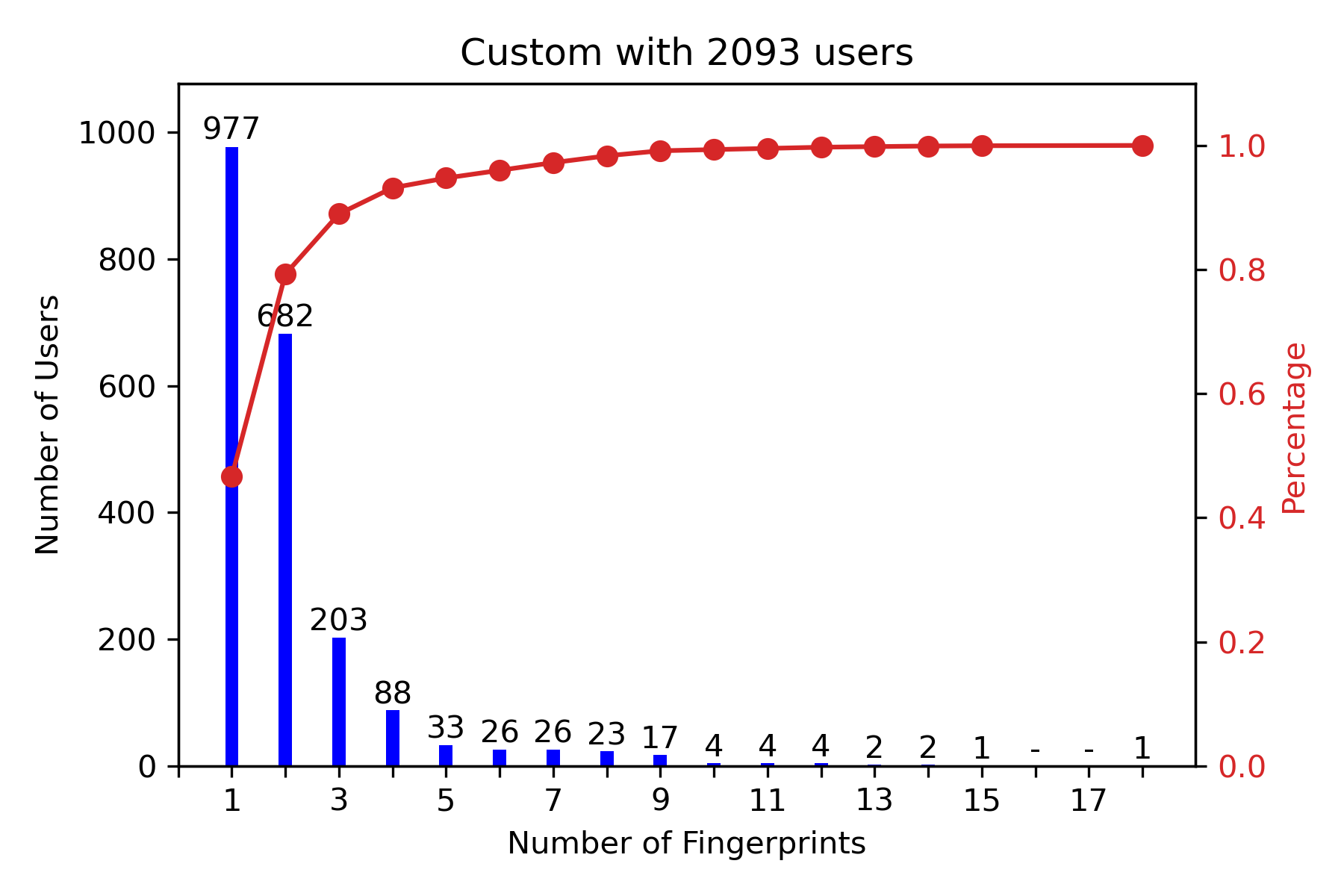}
            \caption{CDFs and Bar plots showing the distribution of number of Custom Signal fingerprints.}
        \end{figure}
        
        \begin{figure}[h]
          \centering
            \includegraphics[width=\columnwidth]{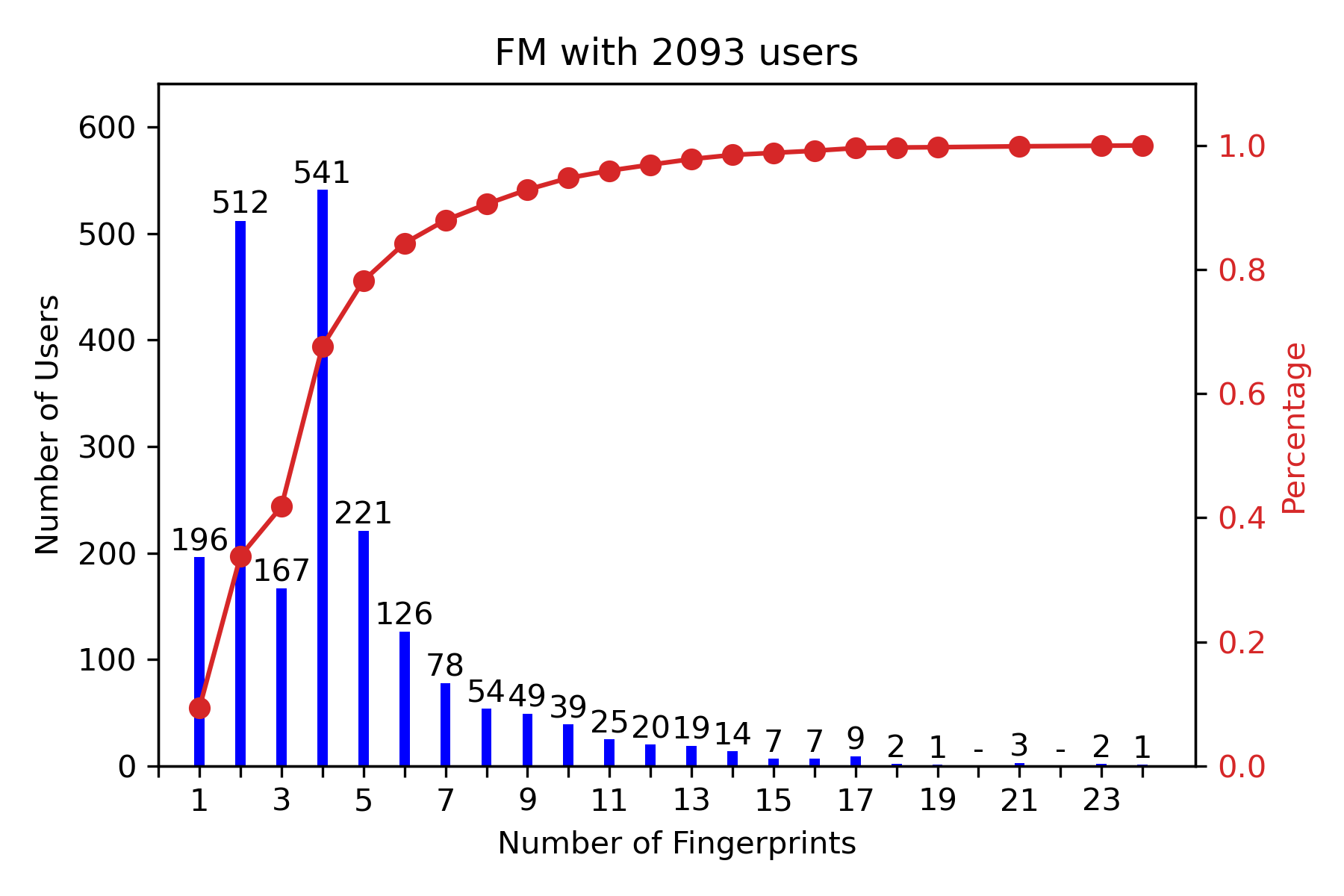}
            \caption{CDFs and Bar plots showing the distribution of number of FM fingerprints.}
        \end{figure}

%
 
 \clearpage
\section{Diversity of fingerprints across smaller subsets}
\label{sec:app_vector_splits}

\begin{table}[h]
\small
\centering
    \begin{tabular}{lcccc}
    \toprule
    {\bf Vectors} & {\bf Distinct} & {\bf Unique} & {\bf Entropy} & $\boldsymbol{e_{norm}}$\\
    \midrule
    Canvas & 146 & 90 & 5.66 & 0.627 \\ 
    \midrule
    Fonts & 227 & 188 & 6.412 & 0.71 \\ 
    \midrule
    UserAgent & 159 & 111 & 5.849 & 0.647 \\ 
    \midrule
    Canvas + Font & 344 & 289 & 7.913 & 0.876 \\ 
    \midrule
    Canvas + Audio & 191 & 130 & 6.106 & 0.676 \\ 
    \midrule
    Canvas + Font + UA & 457 & 416 & 8.707 & 0.964 \\ 
    \midrule
    Canvas + Font + Audio & 365 & 311 & 8.061 & 0.892 \\ 
    \midrule
    Audio FP & 49 & 22 & 2.799 & 0.31 \\ 
    \midrule
    Canvas + Font + UA + Audio & 463 & 425 & 8.737 & 0.967 \\ 
    \bottomrule
    \end{tabular}
    \caption{Diversity of different fingerprinting vectors for first split }
    \label{table:CombinationFingerprints1}
\end{table}

\begin{table}[h]
\small
\centering
    \begin{tabular}{lcccc}
    \toprule
    {\bf Vectors} & {\bf Distinct} & {\bf Unique} & {\bf Entropy} & $\boldsymbol{e_{norm}}$\\
    \midrule
    Canvas & 145 & 87 & 5.701 & 0.631 \\ 
    \midrule
    Fonts & 217 & 170 & 6.426 & 0.712 \\ 
    \midrule
    UserAgent & 171 & 114 & 6.04 & 0.669 \\ 
    \midrule
    Canvas + Font & 347 & 288 & 7.981 & 0.884 \\ 
    \midrule
    Canvas + Audio & 192 & 137 & 6.189 & 0.685 \\ 
    \midrule
    Canvas + Font + UA & 476 & 443 & 8.828 & 0.978 \\ 
    \midrule
    Canvas + Font + Audio & 372 & 315 & 8.174 & 0.905 \\ 
    \midrule
    Audio FP & 44 & 17 & 2.739 & 0.303 \\ 
    \midrule
    Canvas + Font + UA + Audio & 482 & 451 & 8.858 & 0.981 \\ 
    \bottomrule
    \end{tabular}
    \caption{Diversity of different fingerprinting vectors for second split}
    \label{table:CombinationFingerprints2}
\end{table}

\begin{table}[h]
\small
\centering
    \begin{tabular}{lcccc}
    \toprule
    {\bf Vectors} & {\bf Distinct} & {\bf Unique} & {\bf Entropy} & $\boldsymbol{e_{norm}}$\\
    \midrule
    Canvas & 159 & 103 & 5.881 & 0.651 \\ 
    \midrule
    Fonts & 229 & 183 & 6.59 & 0.73 \\ 
    \midrule
    UserAgent & 169 & 118 & 6.06 & 0.671 \\ 
    \midrule
    Canvas + Font & 360 & 307 & 8.035 & 0.89 \\ 
    \midrule
    Canvas + Audio & 213 & 157 & 6.41 & 0.71 \\ 
    \midrule
    Canvas + Font + UA & 474 & 441 & 8.808 & 0.975 \\ 
    \midrule
    Canvas + Font + Audio & 389 & 342 & 8.227 & 0.911 \\ 
    \midrule
    Audio FP & 47 & 27 & 2.809 & 0.311 \\ 
    \midrule
    Canvas + Font + UA + Audio & 477 & 444 & 8.828 & 0.978 \\ 
    \bottomrule
    \end{tabular}
    \caption{Diversity of different fingerprinting vectors for third split}
    \label{table:CombinationFingerprints3}
\end{table}

\begin{table}[h]
\small
\centering
    \begin{tabular}{lcccc}
    \toprule
    {\bf Vectors} & {\bf Distinct} & {\bf Unique} & {\bf Entropy} & $\boldsymbol{e_{norm}}$\\
    \midrule
    Canvas & 149 & 96 & 5.763 & 0.638 \\ 
    \midrule
    Fonts & 223 & 183 & 6.407 & 0.709 \\ 
    \midrule
    UserAgent & 183 & 126 & 6.148 & 0.681 \\ 
    \midrule
    Canvas + Font & 349 & 288 & 7.994 & 0.885 \\ 
    \midrule
    Canvas + Audio & 196 & 134 & 6.233 & 0.69 \\ 
    \midrule
    Canvas + Font + UA & 477 & 444 & 8.831 & 0.978 \\ 
    \midrule
    Canvas + Font + Audio & 378 & 322 & 8.194 & 0.907 \\ 
    \midrule
    Audio FP & 60 & 39 & 2.878 & 0.319 \\ 
    \midrule
    Canvas + Font + UA + Audio & 481 & 451 & 8.847 & 0.98 \\ 
    \bottomrule
    \end{tabular}
    \caption{Diversity of different fingerprinting vectors for fourth split }
    \label{table:CombinationFingerprints4}
\end{table}
\section{Breakdown of Fingerprints by USER AGENTS}
\label{sec:fp_breakdown_by_ua}

\clearpage

\begin{figure*}[hb]
    \centering
    \includegraphics[scale=0.3]{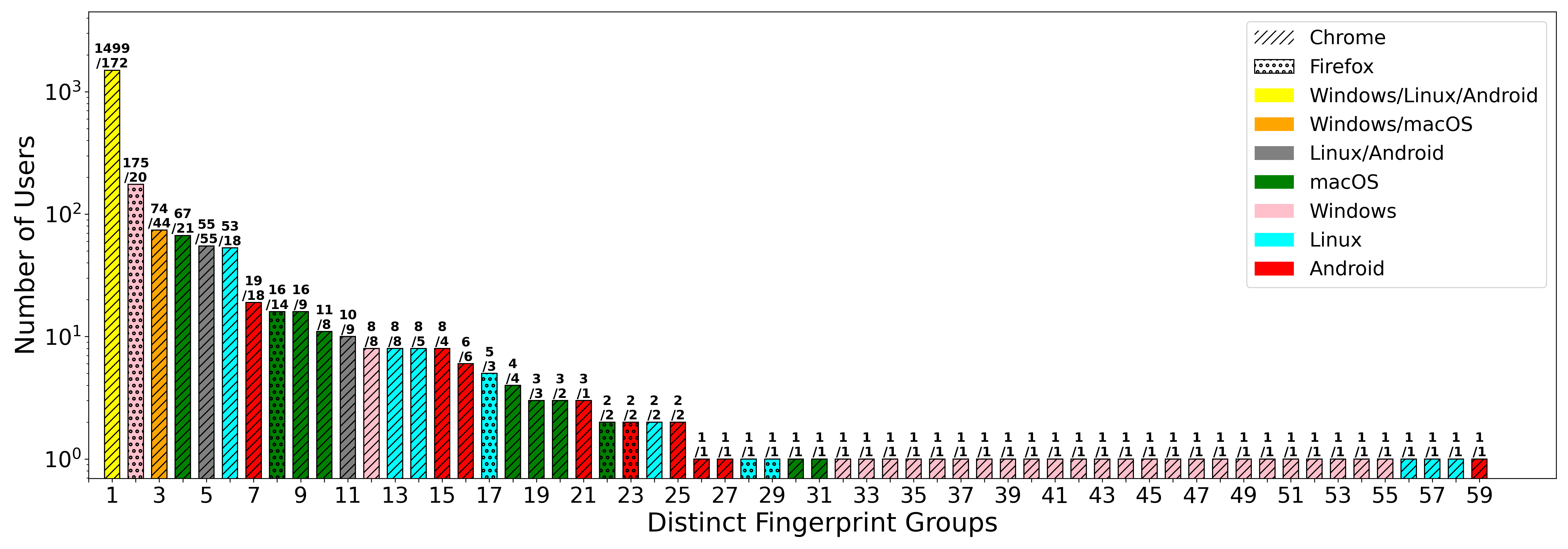}
    \caption{User counts and browser/OS types for Dynamics Compressor (DC) clusters}
\end{figure*}
\begin{figure*}[hb]
    \centering
    \includegraphics[scale=0.3]{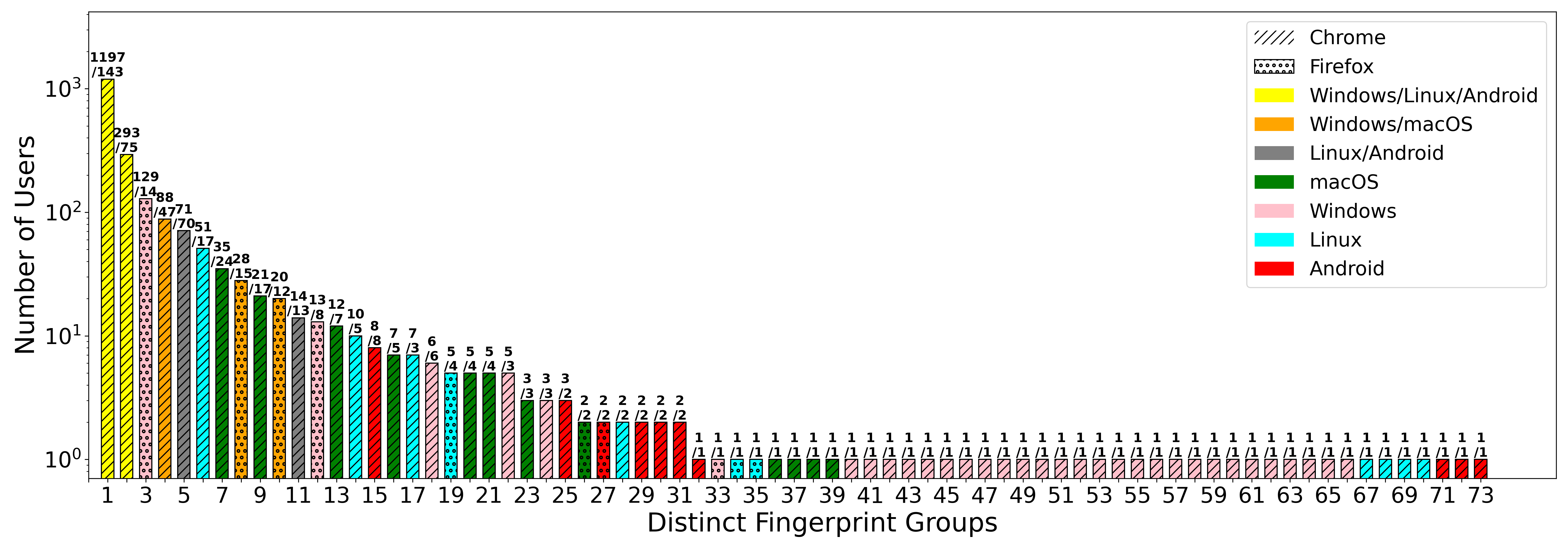}
    \caption{User counts and browser/OS types for FFT clusters}
\end{figure*}
\begin{figure*}[hb]
    \centering
    \includegraphics[scale=0.3]{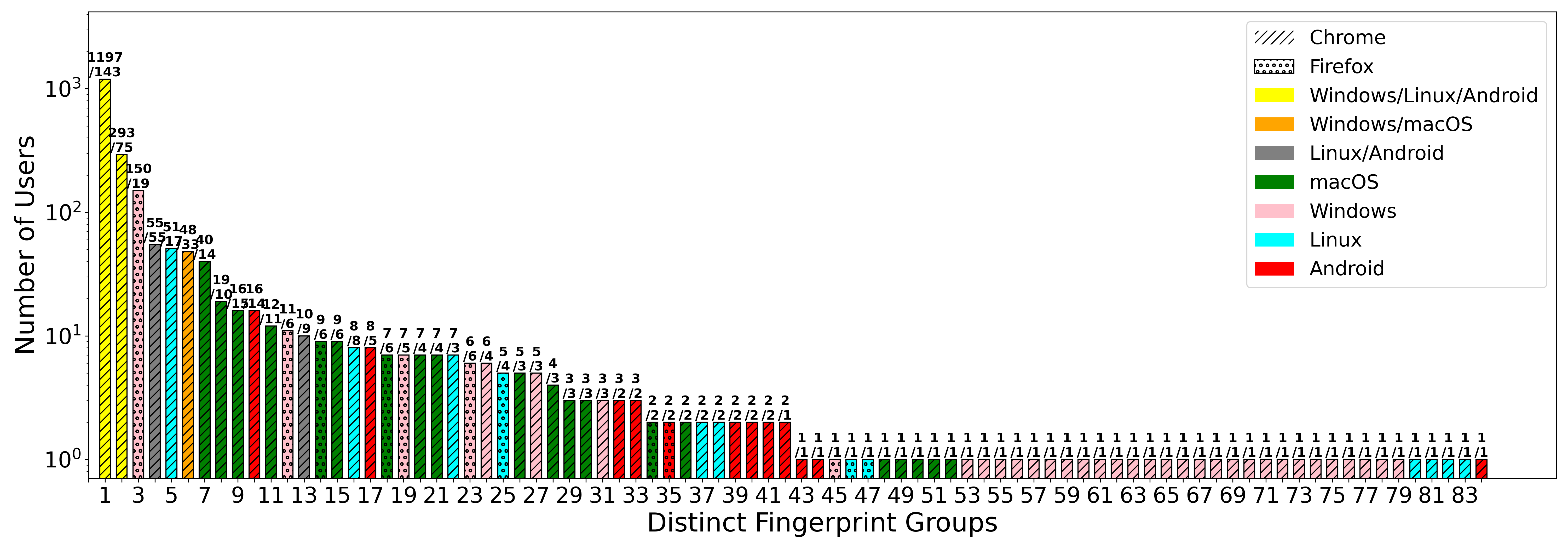}
    \caption{User counts and browser/OS types for Hybrid (DC + FFT) clusters}
\end{figure*}
\begin{figure*}[hb]
    \centering
    \includegraphics[scale=0.3]{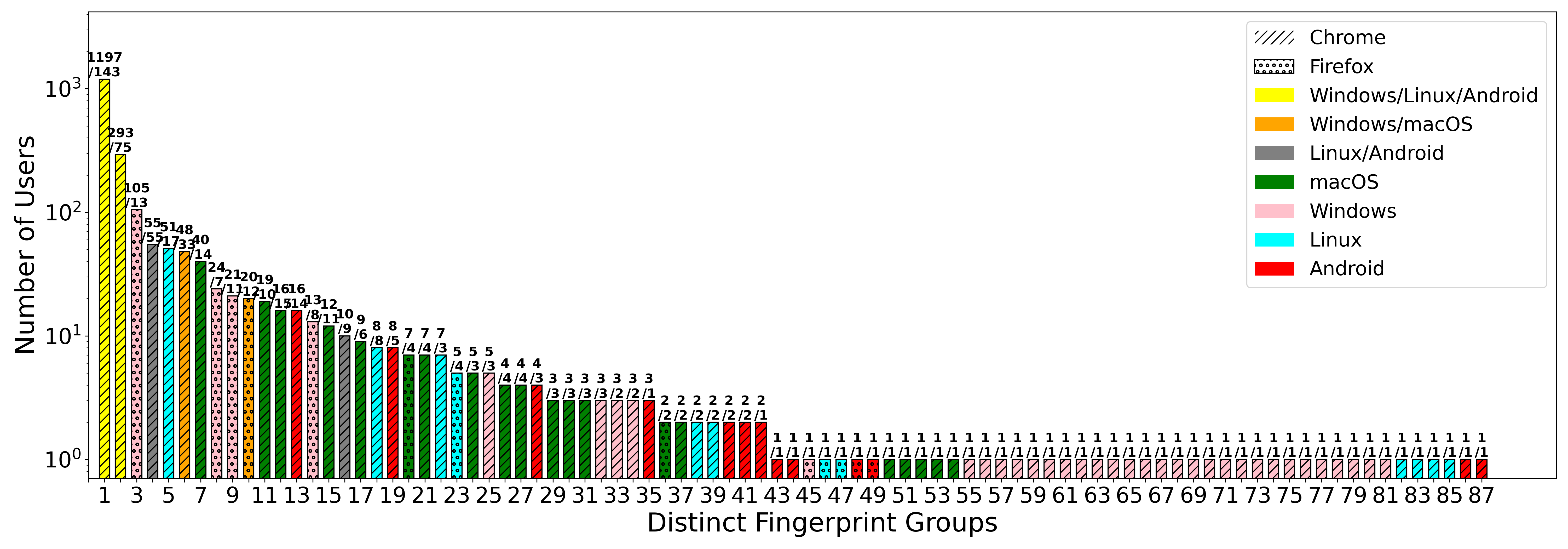}
    \caption{User counts and browser/OS types for Merged Signals clusters}
\end{figure*}
\begin{figure*}[hb]
    \centering
    \includegraphics[scale=0.3]{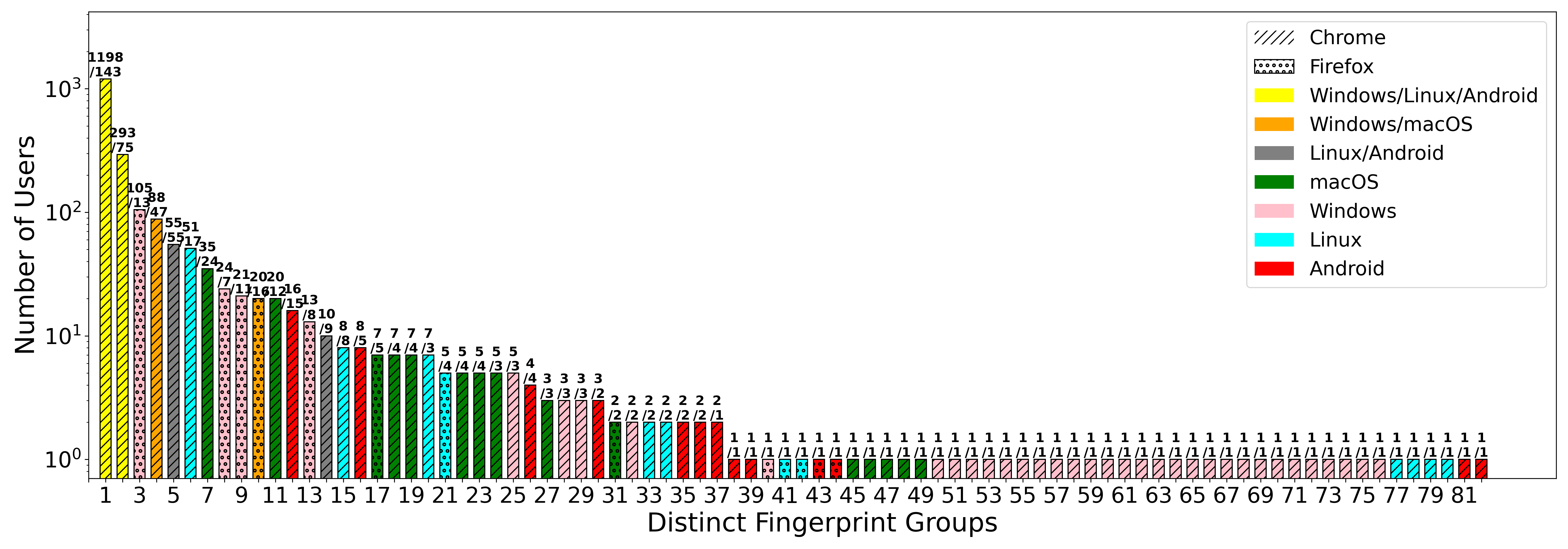}
    \caption{User counts and browser/OS types for Amplitude Modulation (AM) clusters}
\end{figure*}
\begin{figure*}[hb]
    \centering
    \includegraphics[scale=0.3]{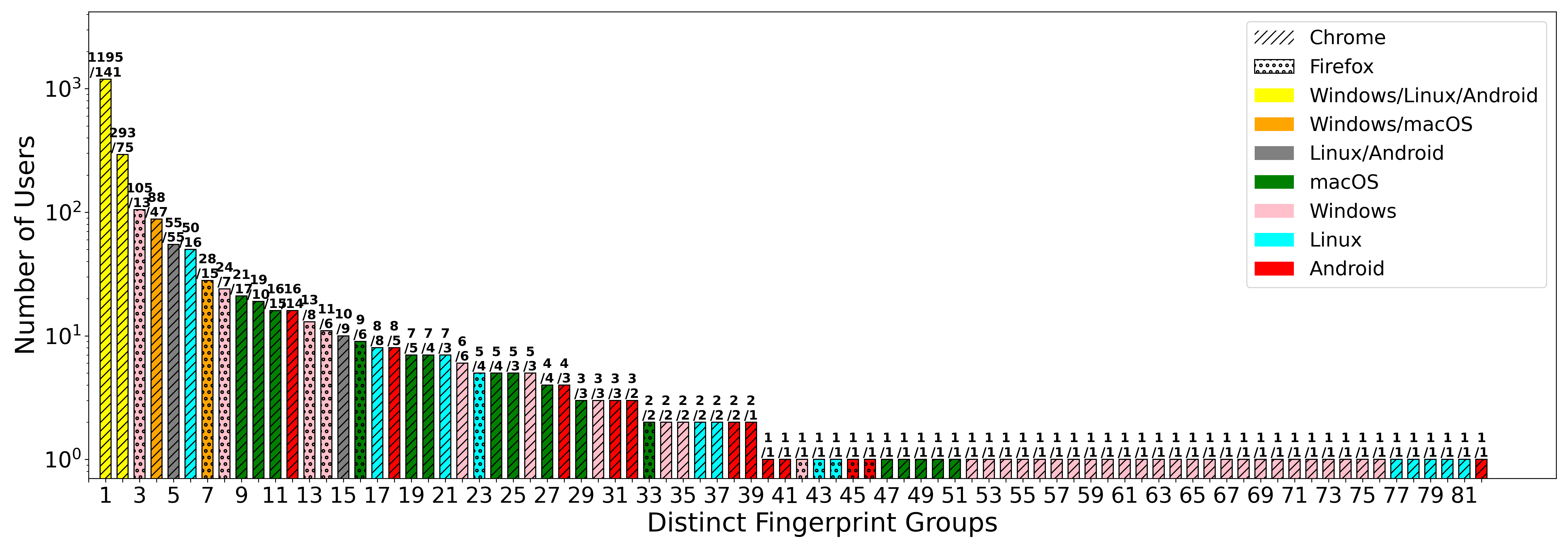}
    \caption{User counts and browser/OS types for Frequency Modulation (FM) clusters}
\end{figure*}

\end{document}